\documentclass[sigconf]{acmart}
\ifdefined\pdfobjcompresslevel \pdfobjcompresslevel=0 \fi
\ifdefined\pdfvariable \pdfvariable objcompresslevel=0 \fi

\AtBeginDocument{%
  }

\setcopyright{cc}
\setcctype{by}

\copyrightyear{2026}
\acmYear{2026}
\setcopyright{cc}
\setcctype{by}
\acmConference[UIST '26]{The 39th Annual ACM Symposium on User Interface Software and Technology}{November 02--05, 2026}{Detroit, MI, USA}
\acmBooktitle{The 39th Annual ACM Symposium on User Interface Software and Technology (UIST '26), November 02--05, 2026, Detroit, MI, USA}
\acmDOI{10.1145/3830398.3830496}
\acmISBN{979-8-4007-2856-3/2026/11}
\usepackage{multirow}
  
\begin{document}
\title{Sighted by Default: Addressing Implicit Vision Assumptions in Real-Time VLM Assistance for BLV Users}
%---------------------------------------------------

\author{Yi Zhao}
\affiliation{%
  \institution{PolyU Shenzhen Research Institute}
  \city{Shenzhen}
  \country{China}
}
\affiliation{%
\institution{COMP, PolyU}
  \city{Hong Kong SAR}
  \country{China}
}
\email{yi-yi-yi.zhao@connect.polyu.hk}

\author{Siqi Wang}
\affiliation{%
  \institution{PolyU Shenzhen Research Institute}
  \city{Shenzhen}
  \country{China}
}
\affiliation{%
\institution{COMP, PolyU}
  \city{Hong Kong SAR}
  \country{China}
}
\email{siqi23.wang@connect.polyu.hk}

\author{Qiqun Geng}
\affiliation{%
  \institution{PolyU Shenzhen Research Institute}
  \city{Shenzhen}
  \country{China}
}
\affiliation{%
  \institution{COMP, PolyU}
  \city{Hong Kong SAR}
  \country{China}
}
\email{qiqun.geng@polyu.edu.hk}

\author{Erxin Yu}
\affiliation{%
  \institution{PolyU Shenzhen Research Institute}
  \city{Shenzhen}
  \country{China}
}
\affiliation{%
\institution{COMP, PolyU}
  \city{Hong Kong SAR}
  \country{China}
}
\email{erxin.yu@outlook.com}

\author{Jing Li}
\authornote{Corresponding author.}
\affiliation{%
  \institution{PolyU Shenzhen Research Institute}
  \city{Shenzhen}
  \country{China}
}
\affiliation{%
\institution{COMP, PolyU}
  \city{Hong Kong SAR}
  \country{China}
}
\email{jing-amelia.li@polyu.edu.hk}

\renewcommand{\shortauthors}{Zhao et al.}

%---------------------------------------------------

\begin{abstract}
Vision-Language Model (VLM)-based assistance is reshaping independence for blind and low-vision (BLV) users, yet current tools fail in dynamic settings. While request-response architectures like BeMyAI impose prohibitive latencies, our formative study (N\,=\,15) reveals that even real-time alternatives like Doubao fail due to a deeper structural problem: \textbf{sighted-default bias}---the implicit assumption that users possess parallel visual access to their surroundings. This bias manifests as verbose, vision-centric narratives that overlook the serial nature of auditory perception, flooding the user's limited cognitive bandwidth with information that is neither timely nor actionable. To address this, we derive three design principles: Continuity, Conciseness, and Calibrated Honesty. We present \textbf{VIA-Agent}, which co-optimizes a specialized cognitive core with a low-latency Real-Time Communication (RTC) architecture for continuous bidirectional streaming. In a within-subjects evaluation (N\,=\,9), VIA-Agent matched Doubao's success rate, significantly reducing mean task time by 21.4\% (91.7s vs. 116.7s) and conversational turns from 5.9 to 4.3 while achieving higher trust.
\end{abstract}

%---------------------------------------------------

\begin{CCSXML}
<ccs2012>
   <concept>
       <concept_id>10003120.10003121</concept_id>
       <concept_desc>Human-centered computing~Human computer interaction (HCI)</concept_desc>
       <concept_significance>500</concept_significance>
   </concept>
   <concept>
       <concept_id>10003120.10011738.10011776</concept_id>
       <concept_desc>Human-centered computing~Accessibility systems and tools</concept_desc>
       <concept_significance>500</concept_significance>
   </concept>
</ccs2012>
\end{CCSXML}

\ccsdesc[500]{Human-centered computing~Human computer interaction (HCI)}
\ccsdesc[500]{Human-centered computing~Accessibility systems and tools}

%---------------------------------------------------
\keywords{blind and low-vision (BLV); real-time assistance;
  Vision-Language Model (VLM); sighted-default bias; cognitive load;
  accessibility}
%---------------------------------------------------

\begin{teaserfigure}
  \includegraphics[width=\textwidth]{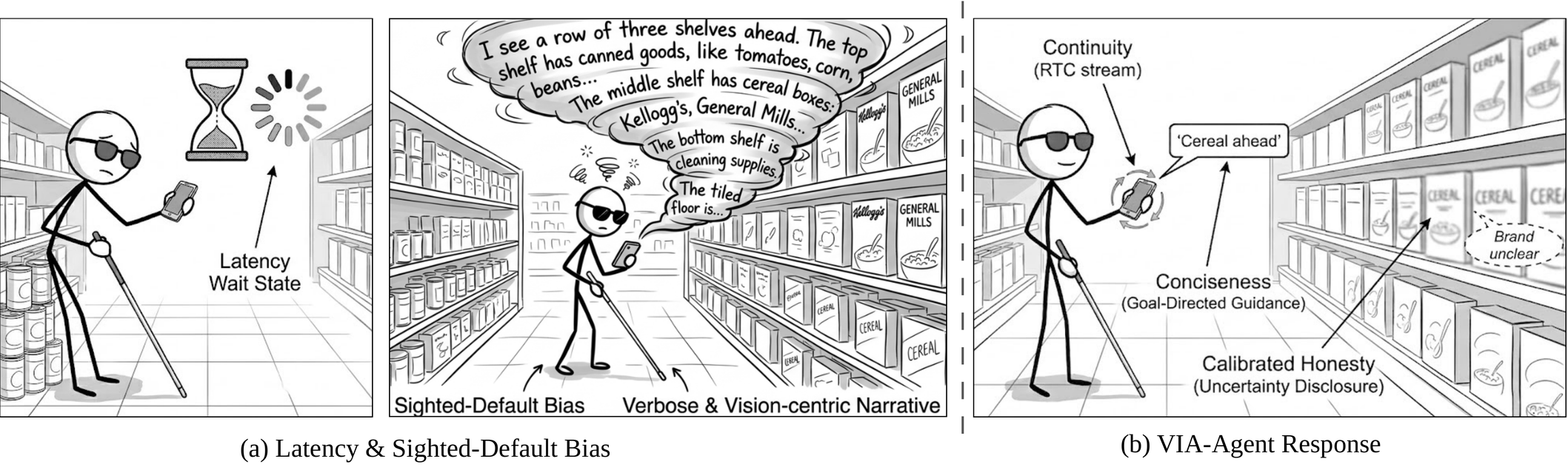}
  \caption{VIA-Agent addresses latency and sighted-default bias. (a) Conventional VLMs involve long wait states or verbose, vision-centric feedback. (b) VIA-Agent ensures Continuity, Conciseness, and Calibrated Honesty.}
    \Description{VIA-Agent addresses latency and sighted-default bias. (a) Conventional VLMs involve long wait states or verbose, vision-centric feedback. (b) VIA-Agent ensures Continuity, Conciseness, and Calibrated Honesty.}
  \label{fig:overall_illustration}
\end{teaserfigure}

\maketitle
\section{Introduction}

% ── HOOK: start from the user, not the technology ──────────────────────────────
For a blind or low-vision (BLV) person in a dynamic environment --- locating
a product in an unfamiliar store, crossing a complex intersection, or tracking
objects in a busy kitchen --- visual assistance must be both \emph{immediate} and
\emph{interpretable}.
Vision-Language Models (VLMs) such as ChatGPT~\cite{openai_gpt5_system_card_2025}
and Gemini~\cite{comanici2025gemini25pushingfrontier} offer transformative potential
for enabling this kind of real-world independence~\cite{zhao2024vialmsurveybenchmarkvisually},
and have spurred dedicated tools including BeMyAI~\cite{bemyai_web_2025} and
general-purpose alternatives like Doubao%
\footnote{Doubao is both a mobile app supporting real-time video chats with AI
and the name of a VLM family.}~\cite{doubao_web_2025}.

% ── PROBLEM STEP 1: the known problem (latency) ────────────────────────────────
Request-response tools like BeMyAI need users to explicitly send a query and
wait for a response, introducing latencies of several seconds --- far too slow for
environments that change faster than the system can
respond~\cite{DBLP:conf/uist/ChangLG24,chang2025probing-gaps-chatgpt-live}.
This limitation drives a shift toward real-time streaming architectures. Tools like Doubao deliver continuous visual narration, seemingly resolving the latency problem.

% ── PROBLEM STEP 2: the deeper structural problem (sighted-default bias) ────────
Yet our formative study with 15 BLV users reveals low latency alone is
insufficient.
Despite real-time descriptions, Doubao's outputs were consistently
rated as overwhelming, disorienting, and difficult to act upon during dynamic
tasks~\cite{pigeon2019cognitive,stepien2019impact}.
Through thematic analysis of participant feedback, we identify a deeper structural
cause: \textbf{sighted-default bias} --- the implicit assumption, embedded in how
current VLMs generate descriptions, that users possess \emph{parallel} visual access
to their environment (Figure~\ref{fig:overall_illustration} (a)).
This assumption drives systems to produce verbose, scene-level narratives optimised
for a sighted observer who can simultaneously see and selectively attend ---
ignoring the \emph{serial} nature of auditory perception.
For BLV users, the result is a flood of information that is neither timely nor
actionable, saturating cognitive bandwidth precisely when clear, goal-directed
guidance is most critical~\cite{chang2025probing-gaps-chatgpt-live}. These observations lead to the overarching research question (\textbf{RQ}) of this paper: \emph{How can real-time VLM assistance address implicit vision assumptions in supporting BLV users?}

% ── SOLUTION: three principles → VIA-Agent ─────────────────────────────────────
To address this research question, we derive three design principles from our formative
findings.
\textbf{Continuity} ensures guidance is proactive and uninterrupted ---
delivered at the moment it becomes actionable, without waiting for an explicit user
query.
\textbf{Conciseness} keeps each response brief and goal-directed, prioritising the
user's immediate objective over comprehensive scene description.
\textbf{Calibrated Honesty} requires the system to disclose uncertainty rather than
produce confident descriptions when visual evidence is ambiguous, preserving user
trust through safe fallbacks~\cite{chang2025probing-gaps-chatgpt-live}.

We present \textbf{VIA-Agent}, a real-time visual assistance system that instantiates
these three principles through two co-optimised components: a \emph{specialised
cognitive core} purpose-built to generate BLV-appropriate, goal-persistent descriptions,
and a \emph{low-latency RTC architecture} enabling continuous bidirectional audio-visual streaming.
Unlike prior tools that adapt sighted-user interfaces to BLV contexts, VIA-Agent is
designed from the ground up around the perceptual and cognitive constraints of BLV
users (Figure~\ref{fig:overall_illustration} (b)). While it builds on a foundation VLM, our contribution is architectural and interaction-centric rather than algorithmic: to our knowledge, VIA-Agent is the first system to deliver BLV-specialised VLM assistance within continuous real-time video live chat.

% ── SOLUTION: three principles → Evaluation ─────────────────────────────────────
We evaluated VIA-Agent in a within-subjects study with 9 BLV participants across navigation and object-retrieval tasks. VIA-Agent reduced mean task time by 21.4\,\% over Doubao (91.7\,s vs.\ 116.7\,s), required fewer conversational turns (4.3 vs.\ 5.9), and significantly outperformed BeMyAI on all NASA-TLX cognitive load dimensions. An ablation study confirms that all three design principles are necessary: removing any single principle measurably degrades performance. This paper makes the following contributions:

$\bullet$~We identify \textbf{sighted-default bias}, a structural failure mode in
current VLM-based assistance, grounded in a formative study with 15 BLV
participants.
This concept reframes the core challenge from a latency problem to a perceptual
mismatch between vision-centric description generation and the serial nature of
auditory perception.

$\bullet$~We derive three design principles --- \textbf{Continuity}, \textbf{Conciseness}, and \textbf{Calibrated Honesty} --- each targeting a manifestation of sighted-default bias in real-time VLM assistance, grounded in formative analysis and validated as necessary through ablation.

$\bullet$~We present \textbf{VIA-Agent}, a real-time visual assistance system
co-optimising a specialised cognitive core with a low-latency RTC architecture for
continuous bidirectional streaming.

$\bullet$~We report results from a \textbf{within-subjects evaluation} (N\,=\,9) demonstrating a 21.4\,\% reduction in task time (91.7\,s vs.\ 116.7\,s) over Doubao and significantly lower cognitive load than BeMyAI on all NASA-TLX dimensions, with ablation confirming each design principle is independently necessary.

\section{Related Works}

\paragraph{\textbf{VLM-based BLV Assistance}}
Research in Visually Impaired Assistance (VIA) supports the daily lives of Blind and Low-Vision (BLV) individuals through technologies that compensate for vision loss~\cite{DBLP:journals/imwut/KianpishehLT19,DBLP:conf/uist/ChangLG24,DBLP:conf/chi/Reinders0M25} and provide non-visual feedback~\cite{DBLP:conf/chi/ClepperMFP25,DBLP:conf/chi/ChenZLWXS25}. Prior work addresses diverse tasks, including navigation~\cite{DBLP:conf/chi/KuribayashiUWMA25, DBLP:conf/chi/ZhaoBBCHMS18,DBLP:conf/chi/SiuSKO0C20,DBLP:journals/imwut/MeinhardtRZECRR24,DBLP:conf/chi/KamikuboKKWKTA25}, shopping~\cite{DBLP:conf/atal/AgrawalNNH23,DBLP:journals/imwut/BolduMZN20}, information access~\cite{DBLP:conf/chi/MoHLDYXY0B25,DBLP:journals/imwut/ZhaoLGLHZ24,DBLP:conf/chi/PereraLCM24,DBLP:conf/chi/Wang0K24}, object manipulation~\cite{DBLP:conf/chi/GuanX024}, household activities~\cite{DBLP:conf/uist/0005TKYPZFTZ24,DBLP:conf/chi/LiLKC24}, social participation~\cite{DBLP:conf/chi/Xie0ZLB024,DBLP:journals/imwut/AhmedKPS18,DBLP:conf/chi/FanTMKPHLSCOLBS25}, and creative work~\cite{DBLP:conf/chi/PandeyOB24,DBLP:conf/chi/KimLP25,DBLP:conf/chi/ClepperMFP25,DBLP:conf/chi/MouallemPMRKKCF25}. These systems are embodied in diverse forms, such as mobile apps~\cite{DBLP:journals/imwut/YangXYLYS21, DBLP:journals/imwut/Ohn-BarGKA18}, wearables~\cite{DBLP:conf/chi/MathisS25,DBLP:journals/imwut/YangXYLYS21,DBLP:journals/imwut/LiuGYYWMS20}, smart glasses~\cite{DBLP:conf/chi/ZhaoBBCHMS18, killough2025vrsight,DBLP:conf/assets/GamageDPLM23}, and embodied agents~\cite{DBLP:conf/chi/HwangJGBL024,DBLP:conf/chi/WeiRGNZOJN25,DBLP:conf/iros/AgrawalWH22}.

Recent systems leverage VLMs like GPT~\cite{openai_gpt5_system_card_2025} and Gemini~\cite{comanici2025gemini25pushingfrontier}, but a gap exists between model capabilities and practical deployment~\cite{chang2025probing-gaps-chatgpt-live, DBLP:conf/assets/GamageDPLM23,DBLP:conf/chi/ShindeM24,DBLP:conf/chi/RanLXFL0L25,DBLP:conf/chi/LuCPEG25,DBLP:conf/chi/NetoHCRHN024,DBLP:conf/chi/JonesGMLM25,DBLP:conf/chi/ChenJMC025,DBLP:conf/chi/India00M025,DBLP:journals/imwut/ZhaoLGLHZ24}. Model-side research improves guidance generation~\cite{yuan2025walkvlmaidvisuallyimpairedpeople,zhao2025lafgrpoinsitunavigationinstruction}, yet specialized models rarely reach end-to-end systems~\cite{zhao2024vialmsurveybenchmarkvisually}. Device-side deployments like BeMyAI~\cite{bemyai_web_2025} and research prototypes~\cite{DBLP:conf/uist/ChangLG24,killough2025vrsight} often use general-purpose VLMs, leading to verbosity and scene description over actionable instruction~\cite{DBLP:journals/imwut/ZhaoLGLHZ24,DBLP:conf/uist/ChangLG24,chang2025probing-gaps-chatgpt-live,meta_ai_glasses_2025}. Most rely on request-response architectures with prohibitive latencies~\cite{DBLP:conf/uist/ChangLG24}. RTC frameworks~\cite{openai_realtime_webrtc,azure_realtime_audio} address latency through continuous streaming, yet sighted-default tendencies persist~\cite{chang2025probing-gaps-chatgpt-live}.
Overall, prior BLV-focused work falls into two groups. The first identifies design principles or diagnoses interaction problems but builds no deployable system~\cite{DBLP:conf/chi/YangSRZ20, chang2025probing-gaps-chatgpt-live}. The second builds task-specific systems that do not jointly support open-ended BLV question-answering and real-time streaming video chat~\cite{DBLP:conf/assets/FroehlichFJTK25, DBLP:conf/chi/WeiAWMOJN26, DBLP:conf/uist/ChangLG24}. VIA-Agent targets both gaps, co-optimizing a VIA-specialized cognitive core with a low-latency RTC embodiment.

\paragraph{\textbf{Cognitive Constraints in BLV Assistance}}
Effective BLV assistance is timely and actionable. Unlike parallel visual processing, BLV users rely on serial auditory perception~\cite{sweller2011cognitive}. Verbosity floods cognitive bandwidth and degrades performance~\cite{DBLP:conf/assets/SharifCWR21,pigeon2019cognitive}, while conciseness reduces this burden~\cite{kosch2023survey}. Prior work views verbosity as an isolated flaw; we trace it to the sighted-default bias in VLM outputs. Our Conciseness principle replaces vision-centric narration with goal-persistent guidance calibrated for serial perception.
\section{Formative Study}
\label{sec:formative}
To ground system design in authentic user needs, we conducted a formative study investigating: (1) persistent challenges for BLV users; (2) systematic failure patterns of current AI assistants; and (3) expectations for next-generation assistive devices.

\subsection{Participants, Procedure, and Data Analysis}
\begin{table*}[t]
  \caption{Participant demographics and assistive technology use. Abbreviations: WC = White cane; SR = Screen reader.}
  \vspace{-1em}
  \label{tab:participants}
  \centering
  \small
  \resizebox{\linewidth}{!}{%
    \begin{tabular}{@{} l l l p{2.5cm} p{2.5cm} p{3.8cm} p{5.2cm} @{}}
      \toprule
      \textbf{ID} & \textbf{Age} & \textbf{Gender} & \textbf{Onset} & \textbf{Vision Status} & \textbf{Assistive Tools} & \textbf{Freq. Used Apps (non-nav)} \\
      \midrule
      P1  & 23 & Male   & Acquired (Adol.) & Light perception & WC; SR; AI tools & TianTan~\cite{tatans}; Be My AI~\cite{bemyai_web_2025} \\ 
      P2  & 25 & Male   & Congenital              & Light perception & WC; SR; AI tools & Envision AI~\cite{envision} 
      \\
      P3  & 24 & Male   & Acquired (Inf.)      & Light perception & WC; SR & TianTan \\
      P4  & 25 & Female & Congenital              & Central vision loss & WC; SR; AI tools & TianTan; ZhengDu~\cite{zdsr}; Doubao~\cite{doubao_web_2025} \\
      P5  & 22 & Male   & Congenital              & Light perception & WC; SR & TianTan  \\
      P6  & 23 & Female & Acquired (Child.)    & Low vision & SR & TianTan  \\
      P7  & 24 & Female & Acquired (Inf.)      & Fully blind & WC; SR; AI tools & iPhone VoiceOver~\cite{apple_voiceover}; Doubao \\
      P8  & 20 & Male   & Acquired (Adol.)  & Fully blind & WC; SR & TianTan; ZhengDu  \\
      P9  & 23 & Female & Acquired (Child.)    & Low vision & WC; SR; Magnifier & Magnifier \\
      P10 & 40 & Male   & Congenital              & Fully blind & WC; SR; AI tools & Doubao \\
      P11 & 42 & Female & Congenital              & Light perception & WC; SR & TianTan \\
      P12 & 40 & Male   & Congenital              & Low vision & Magnifier & Magnifier \\
      P13 & 54 & Male   & Acquired (Adol.)  & Fully blind & WC; SR; AI tools & DianMing~\cite{dianming_sr}; ZhengDu; Doubao \\
      P14 & 37 & Male   & Congenital              & Low vision & SR & iPhone VoiceOver \\
      P15 & 29 & Male   & Acquired (Child.)    & Fully blind & WC; SR; AI tools; remote & Be My Eyes~\cite{bemyai_web_2025}; Doubao \\
      \bottomrule
    \end{tabular}
  }
\end{table*}
We recruited 15 participants (10 men, 5 women; age range 20--54) with diverse visual impairments, including 7 with congenital and 8 with acquired onset, spanning total blindness, light perception, low vision, and central vision loss (Table~\ref{tab:participants}). Most participants relied on white canes or screen readers; 7 had prior experience with AI-based assistive apps (e.g., BeMyAI~\cite{bemyai_web_2025}), while 8 had no prior AI usage. After obtaining IRB approval and informed verbal consent, we conducted 30--45 minute semi-structured remote interviews covering daily challenges, AI-tool frustrations, and future expectations. All data were anonymized and transcribed verbatim for thematic analysis~\cite{braun2006using}. Two researchers independently conducted open coding, then iteratively grouped codes into higher-level themes to inform system design. Findings were attributed to participants based on frequency counts derived from the thematic coding.

\subsection{Findings}
\subsubsection{Daily Challenges Despite Current Practices.} Analysis revealed three core difficulties: (1) \textbf{The last-meter problem.} Most participants (13/15) struggled with the ``last-meter problem,'' where GPS lacks the granularity to locate specific entrances. P4 highlighted the resulting ambiguity: \textit{``The navigation app says the destination has been reached, but there are 20 shops here---which one do I want?''} (2) \textbf{Low latency in dynamic scenarios.} 10/15 participants emphasized that in dynamic environments, delays risk failure or injury. P3 required precision \textit{``to half a second''} for obstacles, while P15 warned: \textit{``Even a little bit of delay can be fatal.''} (3) \textbf{Object identification and information access.} 9/15 participants struggled to locate specific products or granular label details. P2 desired aisle-level guidance to find items without touching them \textit{``one by one.''}

\subsubsection{Frustrations with Current AI Assistants (N=7 with AI Experience).} Participants with prior AI experience reported two primary frustrations: (1) \textbf{A crisis of confidence.} All experienced participants (7/7) reported significant frustration with unreliable model outputs. Dangerous \textbf{hallucinations} were a primary concern. P4 noted the AI \textit{``uses the internet to supplement the parts it can't see clearly, which doesn't match reality, and it doesn't even tell me it can't see clearly.''} This was compounded by \textbf{verbosity}, where the AI failed to understand the user's goal and provided a lengthy, irrelevant narration of the entire scene. Both point to a shared structural assumption: that users can visually filter or verify the AI's output---a presupposition we term \textbf{sighted-default bias} and analyze in \S\ref{sec:synthesis}. (2) \textbf{Latency as a barrier to usability.} Several participants (3/7) identified high latency as a primary barrier, rendering AI systems unusable for dynamic, real-world decision-making. P1 noted the AI's failure to react in real time: \textit{``I said `tell me when you recognize the door,' but I pointed it at the door for a long time and got no reaction.''} The problem is especially acute in mobile situations, where a delayed answer is useless. P15 stressed the wider stakes: \textit{``Even a little bit of delay can be fatal \dots{} I hope it can give real-time feedback.''}

\subsubsection{Expectations and Concerns for Future Devices.} Analysis identified three key themes: (1) \textbf{Primacy of the AI `brain' over the wearable `body'.} Most participants (9/15) welcomed hands-free wearables but prioritized substantive AI intelligence over form factor, dismissing devices that merely mirror smartphone functions as \textit{``gimmicks.''} Value resides in the AI \emph{brain}, not the \emph{body}. Privacy trade-offs were deemed acceptable only if accuracy was demonstrably high; as P5 noted, \textit{``if info is accurate, benefits outweigh the risks.''} (2) \textbf{Goal-oriented, truthful intelligence.} 11/15 participants demanded a shift from generic scene descriptions to task-aligned guidance. P13 emphasized that \textit{``exhaustive narration is useless if it fails to match my needs.''} Trust is paramount; participants required explicit uncertainty disclosure rather than confident hallucinations, with P4 warning against AI that \textit{``supplements the parts it can't see''} without notice. (3) \textbf{Device practicality and interaction.} Nearly all (14/15) required non-occluding audio for environmental awareness, as BLV users \textit{``rely on their ears to perceive the world''} (P1) and frequent alerts can interfere with natural hearing. While vibration suits simple notifications, voice remains essential for complex guidance. Finally, reliability is non-negotiable; P12 noted that \textit{``sudden device failure is unacceptable''} and renders the user \textit{``unsafe.''}
\begin{figure*}[ht]
  \centering
  \includegraphics[width=\linewidth]{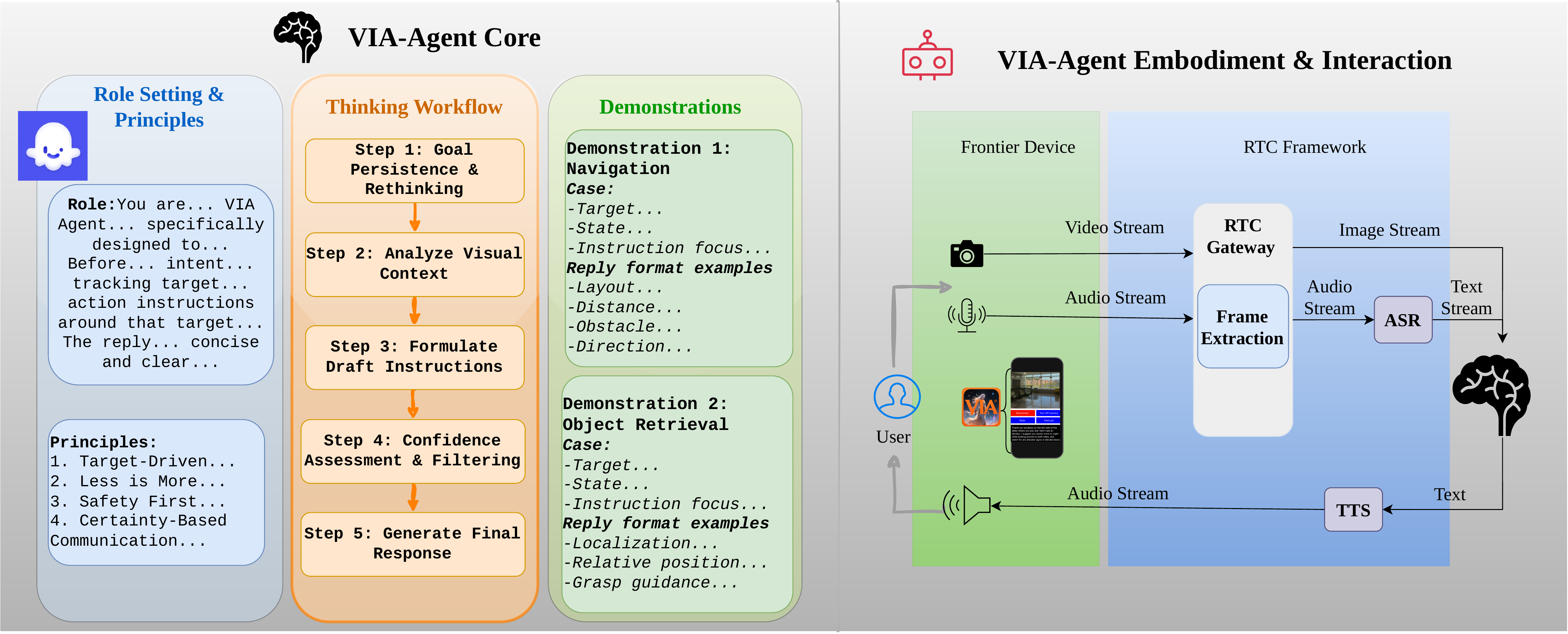}
  \caption{Overview of VIA-Agent architecture, illustrating specialized core and RTC-based embodiment streaming pipeline.}
    \Description{Overview of VIA-Agent architecture, illustrating specialized core and RTC-based embodiment streaming pipeline.}
  \label{fig:system_overview}
\end{figure*}
\subsection{Sighted-Default Bias and Design Principles}
\label{sec:synthesis}
Across our findings, a structural pattern emerges: current AI assistants implicitly assume users possess parallel visual access to their environment. We term this \textbf{sighted-default bias}. It manifests as: (1) \textit{verbosity} that overwhelms serial auditory processing; (2) \textit{latency} that assumes users can safely pause; and (3) \textit{overconfidence} that assumes users can visually verify claims. This is not merely an interface flaw but a structural mismatch between vision-centric AI logic and the audio-first perceptual ecology of BLV users. Findings yield three design principles: (1) \textbf{Continuity:} Abandoning discrete request-response for continuous streaming guidance. In dynamic settings, low latency is a safety constraint, not a performance trade-off. (2) \textbf{Conciseness:} Replacing exhaustive descriptions with goal-directed, brief responses calibrated to the serial nature of auditory processing. (3) \textbf{Calibrated Honesty:} Bounding outputs to verifiable evidence. Systems must disclose uncertainty rather than confabulating, treating overconfident output as a safety hazard.

\section{Prototype Development: The VIA-Agent}
Guided by the three proposed design principles, we systematically developed the VIA-Agent (Figure~\ref{fig:system_overview}): the VIA-Agent Core operationalizes both Conciseness and Calibrated Honesty via goal-persistent reasoning and confidence filtering, while the VIA-Agent Embodiment ensures Continuity through an RTC streaming pipeline.

\begin{figure}[t]
  \centering
  \includegraphics[width=\columnwidth]{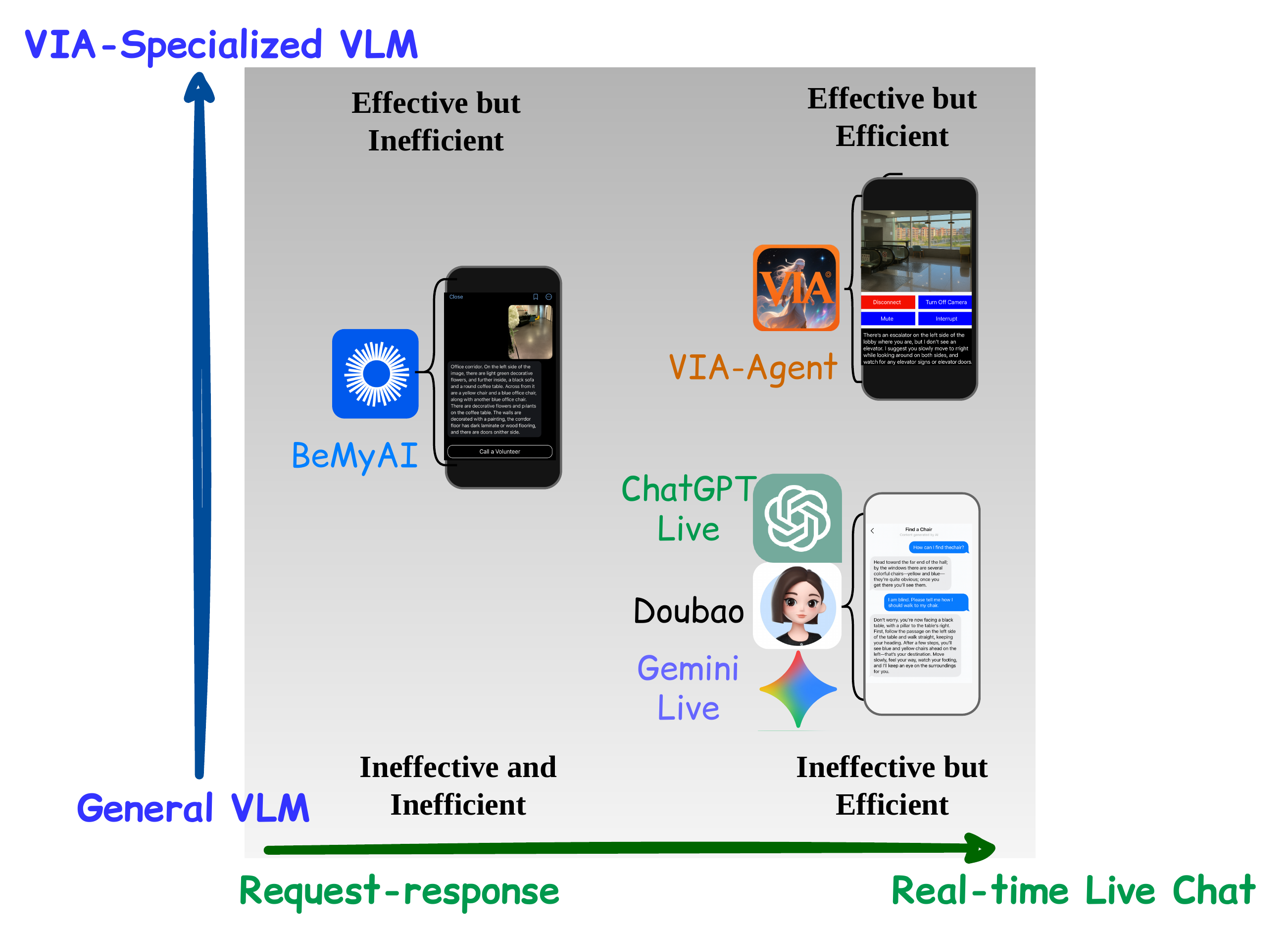}
\caption{Two-axis positioning of visual-assistance approaches: interaction paradigm against model specialization.}
\Description{Two-axis positioning of visual-assistance approaches: interaction paradigm against model specialization.}
  \label{fig:system_comp}
\end{figure}

\textbf{System Comparison.} Figure~\ref{fig:system_comp} positions VIA-Agent against existing systems along two dimensions: interaction mode and VIA specialization. Request--response tools (e.g. BeMyAI) provide BLV-oriented assistance but require users to stop, capture, and wait. General-purpose streaming systems, such as Doubao, Gemini Live, and ChatGPT Live, support real-time live chat but are not designed around BLV users' serial auditory perception or task-specific guidance needs. VIA-Agent occupies the intersection, combining real-time live chat with a VIA-specialized cognitive core.

\subsection{The VIA-Agent Core}
\label{sec:agent_core}

The VIA-Agent Core instantiates the \textbf{Conciseness} and \textbf{Calibrated Honesty} principles through a purpose-built VLM cognitive architecture. Off-the-shelf VLMs — trained primarily on sighted-user corpora — generate scene-level narratives optimised for visual parallel processing, directly manifesting the sighted-default bias identified in our formative study. This bias manifests in two critical failure modes: \textit{task drift} (losing track of the user's goal amid verbose scene descriptions) and \textit{overconfident hallucinations} (generating unverifiable claims without epistemic disclosure). The Core addresses these through principle-driven architectural constraints.

We employ a VLM with explicit chain-of-thought (COT) reasoning~\cite{DBLP:conf/nips/Wei0SBIXCLZ22} to perform a structured, step-by-step analysis before generating a response. As depicted in Figure~\ref{fig:system_overview} left, the cognitive architecture is defined by three components: a foundational \textit{Role Setting \& Principles}, a multi-step \textit{Thinking Workflow}, and task-specific \textit{Demonstrations} provided as in-context learning (ICL) ~\cite{DBLP:conf/emnlp/Dong0DZMLXX0C0S24}. For our implementation, we utilised \texttt{Doubao-1.6-thinking-250715} as the base VLM, which provides native COT reasoning capabilities.
\begin{figure}[htbp]
  \centering
  \includegraphics[width=0.9\linewidth]{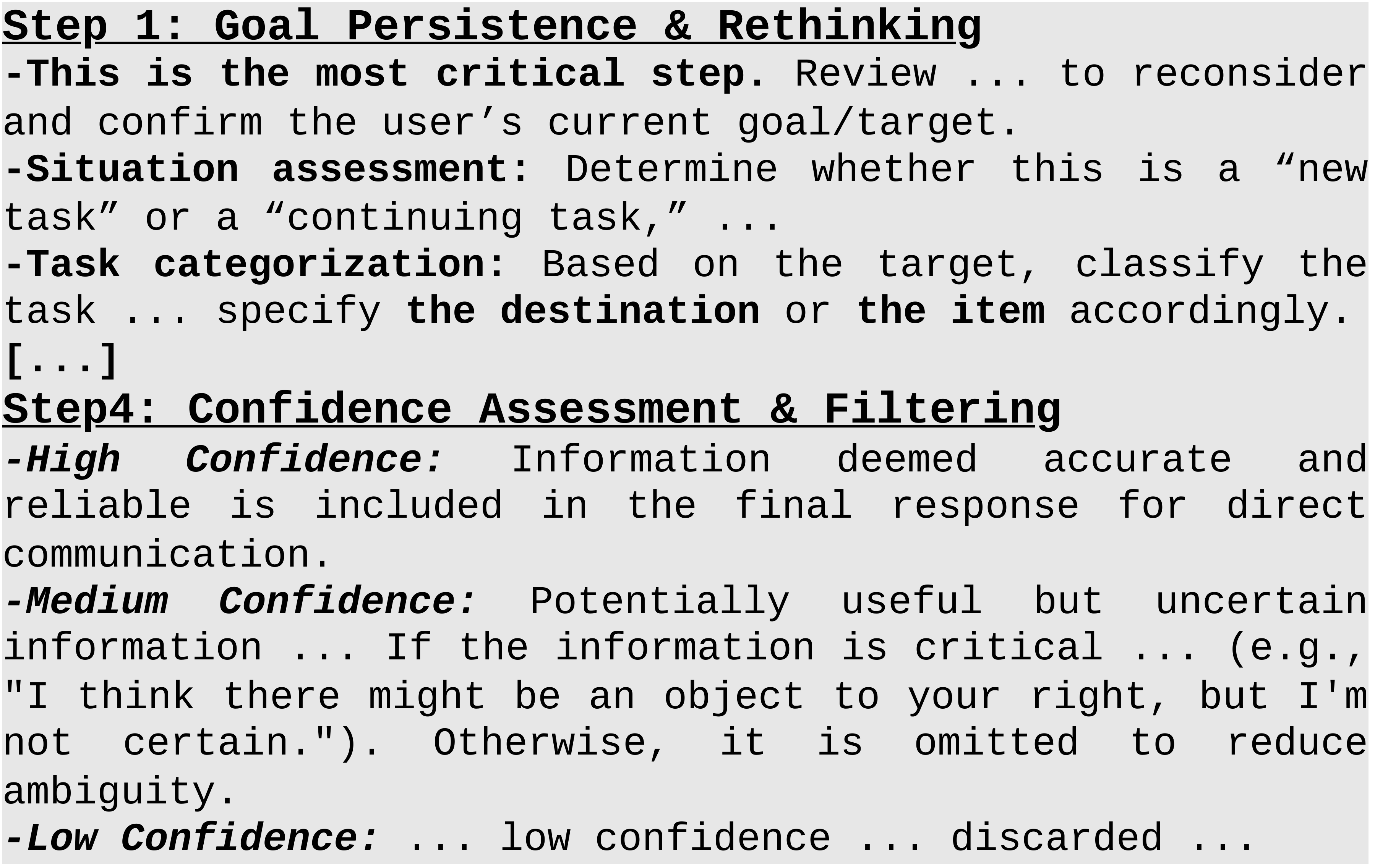}
  \caption{Procedural logic of VIA-Agent Core.}
  \Description{Procedural logic of VIA-Agent Core.}
  \vspace{-1em}
  \label{fig:agent_core_para}
\end{figure}
\subsubsection{Operationalising Conciseness.}
The \textbf{Conciseness} principle requires that the agent deliver only goal-relevant information, eliminating the verbose scene-level narration characteristic of sighted-default systems. Our formative study identified two manifestations of verbosity-induced failure: \textit{task drift} (the AI loses track of the user's goal amid irrelevant descriptions) and \textit{information overload} (excessive output overwhelms serial auditory processing). We operationalise Conciseness through two mechanisms:

\textbf{Goal-Persistent Design.}
Task drift is a direct symptom of sighted-default bias: a sighted user can visually re-anchor to their goal at any moment, while a BLV user relying on serial audio cannot. To ensure the agent remains persistently focused on the user's goal, we implemented three architectural constraints. \textbf{(1) Mandatory Goal Re-evaluation}: as the first step in its Thinking Workflow (\textit{Goal Persistence \& Rethinking}), the agent must explicitly re-state the user's current goal before processing any new input, preventing distraction by irrelevant visual stimuli. \textbf{(2) Short-Term Conversational History}: the agent maintains a sliding context window of the two most recent interaction pairs (\texttt{Number of context rounds = 2}), providing sufficient context for follow-up instructions while preventing irrelevant history from obscuring the current goal. \textbf{(3) Persistent Session Memory}: we enabled long-term memory with a persistence of one day, ensuring the agent can resume tasks across interrupted sessions without treating each restart as a new, contextless interaction.

\textbf{Enforced Brevity.}
To counteract the model's tendency for verbose scene-level narration, we apply a hard constraint limiting responses to a maximum of 128 tokens (\texttt{Response max length = 128}). This threshold is informed by prior work on in-situ navigation~\cite{zhao2025lafgrpoinsitunavigationinstruction}, which shows effective navigation instructions are typically concise ($<$100 tokens); our slightly higher limit provides a buffer for complete responses without enabling full scene narration.

\subsubsection{Operationalising Calibrated Honesty.}

The \textbf{Calibrated Honesty} principle requires that the agent disclose uncertainty rather than generate overconfident claims that assume the user can visually verify their accuracy. Our formative study revealed that all participants with AI experience (7/7) reported frustration with hallucinations delivered without epistemic disclosure — a system behaving as if the user possesses parallel visual access to fact-check its output. This is a structural manifestation of sighted-default bias.

To operationalise Calibrated Honesty, the agent executes a rigorous \textbf{Multi-Level Confidence Filtering} mechanism as Step~4 of its Thinking Workflow (Figure~\ref{fig:agent_core_para}). The agent scrutinises every internally generated claim (e.g., object identity, spatial distance, navigational instruction) and assigns one of three confidence levels. A strict filtering rule governs each level: \textit{high-confidence} claims are delivered directly; \textit{uncertain} claims are hedged with explicit disclosure (e.g., ``I think this might ...''); \textit{low-confidence} claims are withheld entirely in favour of a safe fallback response (e.g., ``I cannot clearly identify this''). This counters the hallucination-without-disclosure pattern reported by formative study participants, preventing the agent from presenting unverifiable information as fact.

Together, the Conciseness and Calibrated Honesty mechanisms ensure the Core generates brief, goal-relevant, and epistemically honest guidance — eliminating the information flood and false confidence constituting sighted-default bias during response-generation.

\subsubsection{Human-in-the-Loop Iterative Refinement.}
\label{sec:hitl}
\begin{figure}[t]
  \centering
  \includegraphics[width=0.85\linewidth]{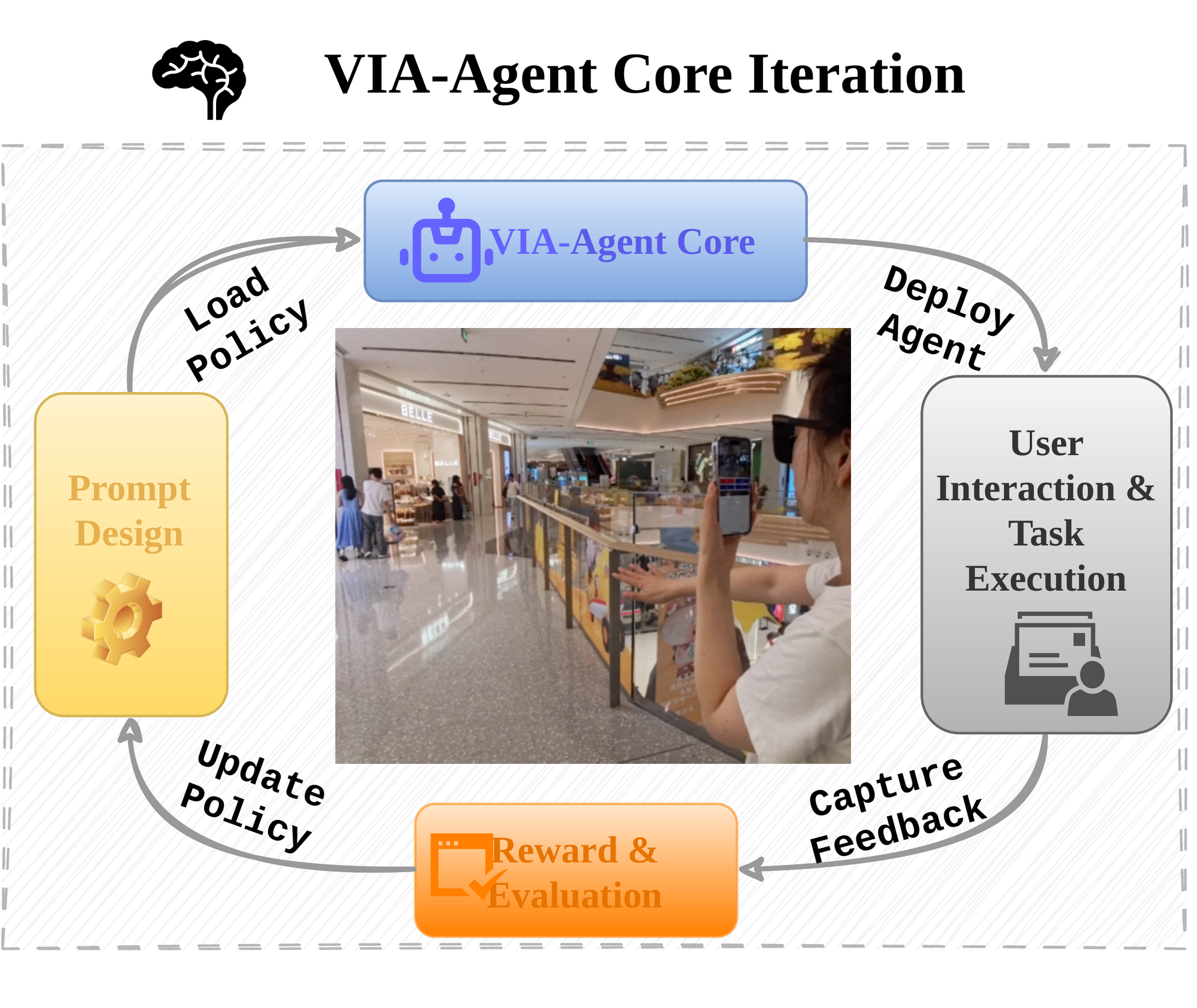}
  \caption{The iterative refinement process for operationalizing design principles into concrete VIA-Agent behaviors.}
\Description{The iterative refinement process for operationalizing design principles into concrete VIA-Agent behaviors.}
  \vspace{-1em}
  \label{fig:iteration}
\end{figure}
Design principles provide the structural constraints; iterative refinement with real users operationalises those constraints into concrete, tunable agent behaviours. As illustrated in Figure~\ref{fig:iteration}, we employed a continuous refinement process with 2 BLV participants across navigation and object-retrieval tasks. Each cycle proceeded through four stages: \textbf{(1) Policy Design}, formulating the agent's operational policy (system prompts, few-shot demonstrations, and parameter constraints) grounded in the three design principles; \textbf{(2) Situated Deployment}, loading the policy into the VIA-Agent Core for real-world use; \textbf{(3) Feedback Capture}, assessing quantitative metrics (task completion time, interaction turn count, success rate) and qualitative feedback as a rapid diagnostic check to identify principle violations; and \textbf{(4) Policy Update}, refining system prompts and demonstrations based on observed violations. This process transformed the design principles from aspirational goals into measurable, falsifiable properties of the agent’s behaviour. An iteration concluded when outputs consistently aligned with the three principles across evaluated scenarios, rather than prioritizing satisfaction scores.

\subsection{The VIA-Agent Embodiment}
\label{sec:agent_body}
The VIA-Agent Embodiment instantiates \textbf{Continuity} through a low-latency streaming architecture. Our formative study identified response latency as a critical barrier for BLV users in dynamic environments: participants emphasised that delayed guidance is useless — and potentially dangerous — when surroundings change faster than the system can respond (§\ref{sec:formative}). Continuity requires that guidance be delivered \textit{at the moment it becomes actionable}, ruling out any request-response interaction paradigm despite optimization.

\subsubsection{Request-Response Architecture: A Controlled Baseline.}
\label{sec: MCP_embodiment}

\begin{figure}[htbp]
  \centering
  \includegraphics[width=0.8\linewidth]{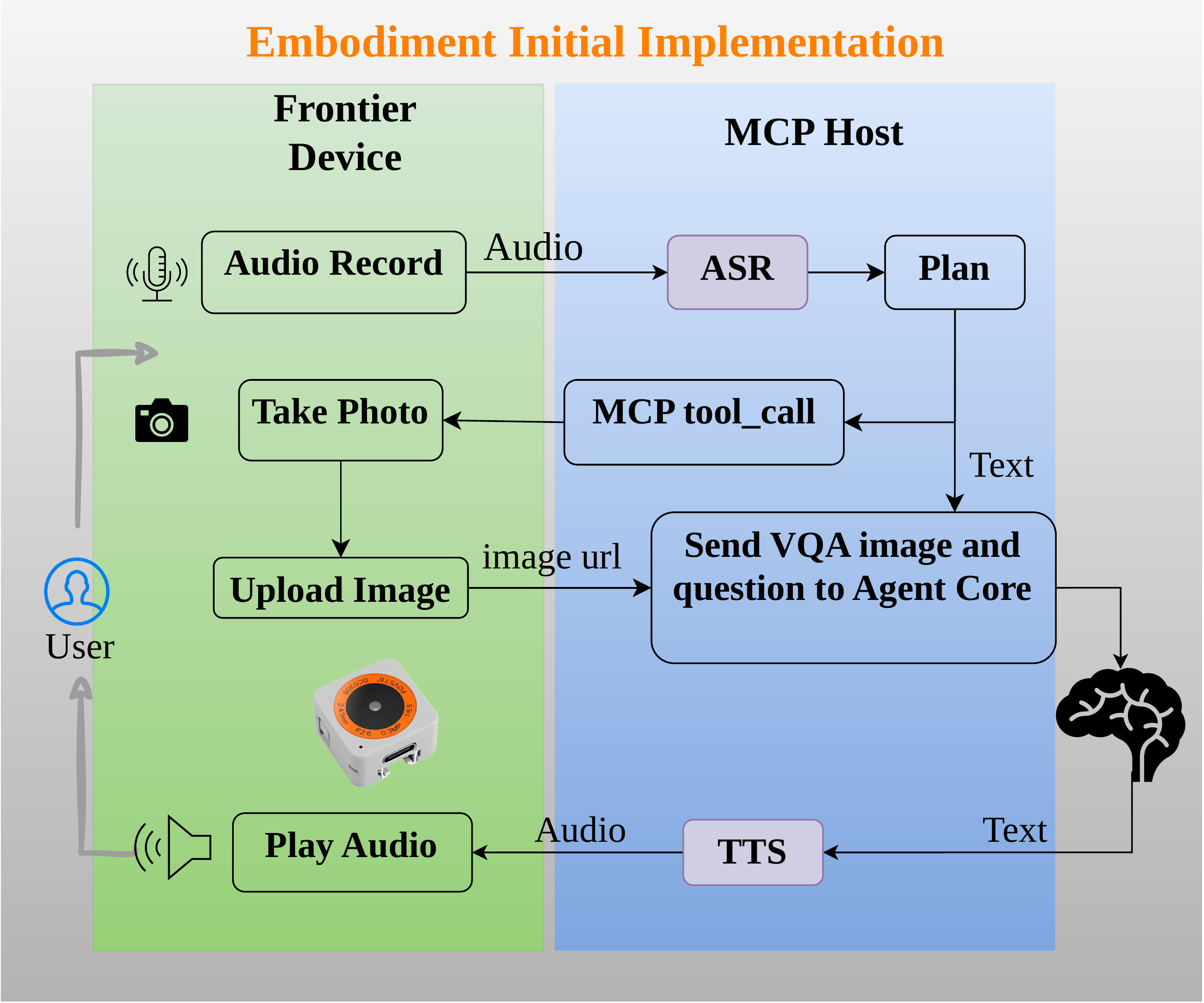}
  \caption{Initial VIA-Agent Embodiment Architecture.}
  \Description{Initial VIA-Agent Embodiment Architecture.}
  \vspace{-1em}
  \label{fig:initial_mcp}
\end{figure}

To validate this architectural constraint empirically, we first prototyped the embodiment on a Model Context Protocol (MCP) request-response framework as a controlled baseline (Figure~\ref{fig:initial_mcp}). This design held the VIA-Agent Core constant while varying only the interaction architecture, allowing us to isolate the effect of the request-response paradigm on Continuity.

Performance profiling revealed per-turn end-to-end latency consistently exceeding 15 seconds under real-world conditions. This latency was not merely a performance shortcoming, it was a \textit{structural violation} of the Continuity principle. The MCP's request-response model requires users to pause physical activity to initiate a query and await a response, replicating the stop-and-go interaction pattern our formative study identified as incompatible with dynamic BLV tasks (§\ref{sec:formative}). Standard optimization cannot resolve this mismatch: the architecture itself assumes a user who can afford to wait.

The MCP prototype thus served as empirical evidence for the Continuity principle's necessity: even with the VIA-Agent Core's cognitive capabilities held constant, a request-response architecture rendered the system unusable for real-time BLV assistance.

\subsubsection{Operationalising Continuity through RTC Streaming.}

To instantiate \textbf{Continuity}, we re-architected the embodiment on a Real-Time Communication (RTC) streaming framework (Figure~\ref{fig:system_overview} right). This shift from discrete request-response pipeline to \textit{continuous, bidirectional streaming} is not an optimization but the minimal architectural change necessary to satisfy the Continuity constraint.

\textbf{RTC pipeline.}
As illustrated in Figure~\ref{fig:system_overview}, the mobile application establishes a persistent media stream with the cloud-based agent. The phone continuously streams video from its camera and audio from its microphone to an RTC Gateway. The video stream undergoes continuous frame extraction to produce an image stream; the audio stream is concurrently transcribed by a cloud Automatic Speech Recognition (ASR) module into a text stream. Both streams are fed in parallel to the VIA-Agent Core for analysis. The Core processes the continuous inputs to generate incremental textual responses; actionable guidance is forwarded to a Text-to-Speech (TTS) service, which synthesises and streams audio back to the user — maintaining \textbf{Continuity} throughout the interaction cycle.

\begin{figure}[t]
  \centering
  \includegraphics[width=\linewidth]{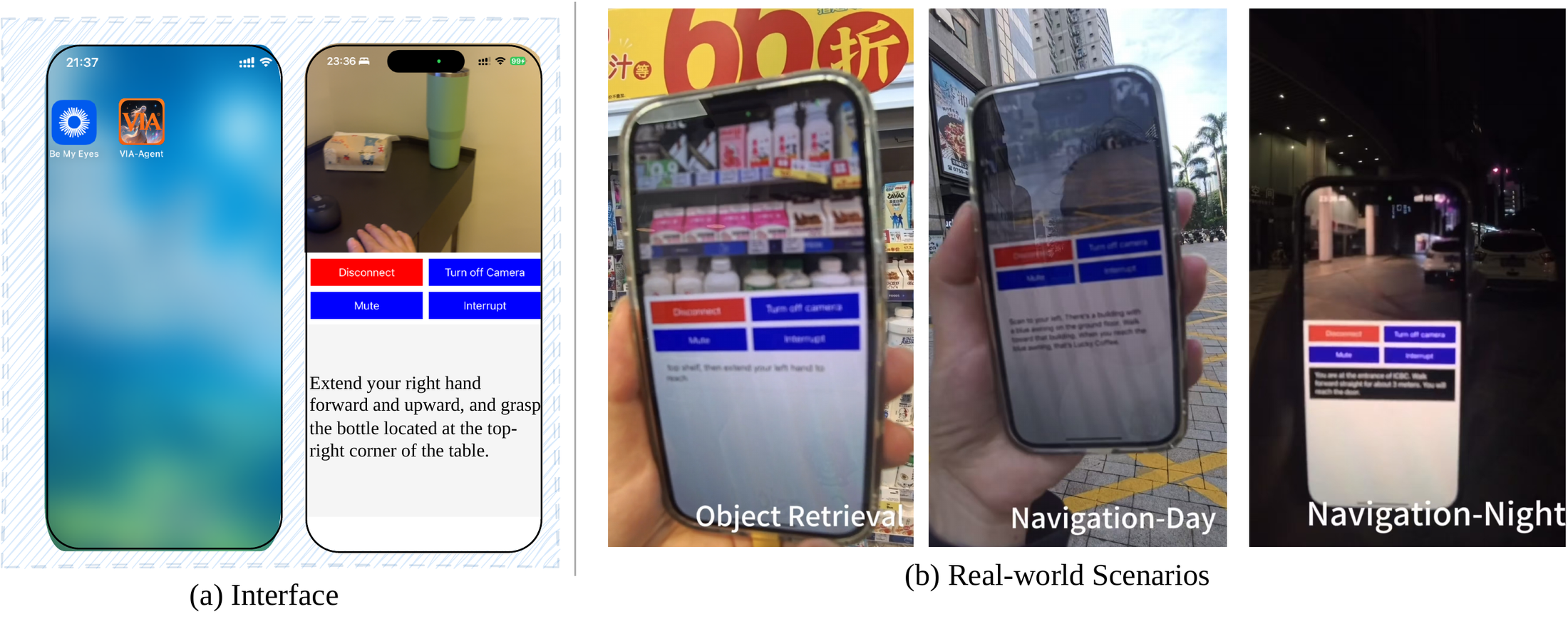}
  \caption{VIA-Agent mobile embodiment and deployment.}
  \Description{VIA-Agent mobile embodiment and deployment.}
  \vspace{-1.5em}
  \label{fig:app_overview}
\end{figure}

\textbf{Platform rationale.}
Our decision to implement the RTC embodiment as a mobile application rather than a wearable device was a principled trade-off grounded in three considerations. First, BLV users are already proficient smartphone users — most participants in our formative study relied on screen readers daily — making the smartphone \textbf{a zero-adoption-barrier platform} (Figure~\ref{fig:app_overview}). Second, low-power microcontrollers (e.g., ESP32-S3) lack integrated support for multimodal video RTC, while more capable alternatives (e.g., ESP32-P4) have power and form-factor profiles inconsistent with all-day wearable use. Third, few AI glasses available during development exposed the bidirectional
  multimodal streaming that Continuity requires. Even on more capable head-worn
  hardware, such as the Rokid AR Station~\cite{rokid_arstudio_web_2026}, we found
  additional barriers that made the platform unsuitable. Our attempt to prototype VIA-Agent on this AI-glasses platform encountered two
  blockers: the RTC-compatible camera path provided
  image quality too low for reliable recognition, and menu navigation depended on
  ray-casting, which is unusable without vision. We therefore leveraged smartphones to prioritise the Continuity requirement, treating wearable form factor as a secondary constraint pending hardware maturation.

\section{Evaluation}
\label{sec:evaluation}
We conducted a three-part evaluation: (1) user study with nine BLV participants comparing VIA-Agent against two real-world baselines; (2) technical evaluation analyzing performance metrics; and (3) ablation study isolating contributions of each design principle. Appendix~\ref{sec:appendix_in_use} illustrates participants using VIA-Agent across representative navigation and object-retrieval tasks.

\subsection{User Evaluation}
\label{sec:user_eval}
\subsubsection{Participants}
We recruited nine participants (P1--P9) with visual impairments via local
blindness advocacy groups. The cohort comprised four males and five females
(mean age = $35.7$, $SD = 14.1$; range $= 23$--$67$). Vision levels
ranged from total blindness to varying degrees of low vision. All participants were experienced smartphone users. Full demographics are provided in Table~\ref{tab:evaluation_participants}.
\begin{table}[h!]
  \centering
  \caption{Demographics of user evaluation participants (P1--P9).
           WC\,=\,White Cane; SR\,=\,Screen Reader.}
  \label{tab:evaluation_participants}
  \resizebox{\columnwidth}{!}{%
    \begin{tabular}{@{} l l l l l l @{}}
      \toprule
      \textbf{ID} & \textbf{Age} & \textbf{Gender} & \textbf{Onset}
        & \textbf{Vision Status} & \textbf{Assistive Tools} \\
      \midrule
      P1 & 29 & Male   & Acquired (Child.) & Fully blind      & WC; SR; AI tools \\
      P2 & 31 & Female & Congenital        & Low vision       & SR; Magnifier \\
      P3 & 42 & Female & Congenital        & Light perception & WC; SR; AI tools \\
      P4 & 47 & Male   & Acquired (Adol.)  & Light perception & WC; SR \\
      P5 & 32 & Female & Acquired (Adol.)  & Low vision       & WC; SR \\
      P6 & 25 & Female & Acquired (Child.) & Low vision       & SR; Magnifier \\
      P7 & 26 & Male   & Congenital        & Low vision       & SR; AI tools \\
      P8 & 67 & Male   & Acquired (Adol.)  & Light perception & WC; SR \\
      P9 & 23 & Female & Acquired (Adol.)  & Light perception & WC; SR \\
      \bottomrule
    \end{tabular}%
  }
\end{table}

Although $N=9$ is modest, it is consistent with comparable BLV studies involving real-world tasks~\cite{killough2025vrsight, huh2025vid2coach}.  Our within-subjects design maximized statistical power, yielding 243 total trials and achieving adequate sensitivity to detect significant effects. The study was IRB-approved; all participants provided informed consent prior to participation and received \$25 as compensation.
% ── Apparatus and Baselines ─────────────────────────────────────────────
\subsubsection{Apparatus and Baselines}
All evaluations were conducted on an iPhone~14~Pro, connected to either a stable indoor Wi-Fi network or an outdoor 5G cellular network. The evaluated systems were:

$\bullet$~\textbf{VIA-Agent (Our Prototype).} Integrates RTC streaming with a BLV-specialized VLM core, operationalizing all three design principles (Continuity, Conciseness, Calibrated Honesty).

$\bullet$~\textbf{BeMyAI~\cite{bemyai_web_2025} (No-Continuity Baseline).} A widely-used assistive app (v6.10.2) requiring manual photo capture for static image captioning and VQA. Lacks persistent streaming and task-specialized cognitive core, serving as the request-response baseline.

$\bullet$~\textbf{Doubao~\cite{doubao_web_2025} (Continuity-Only Baseline).} A general-purpose conversational AI (v10.6.0) providing real-time video streaming via RTC. Satisfies Continuity but lacks BLV-specialized optimization, thus failing to provide Conciseness or Calibrated Honesty.

\textit{Baseline selection rationale.}
Following prior work~\cite{chang2025probing-gaps-chatgpt-live}, we use BeMyAI as
  a representative and widely used photo-based request-response baseline, a
  category that also includes Seeing AI~\cite{seeingai_web_2026} and
  Lookout~\cite{lookout_play_2026}. Doubao represents real-time bidirectional video assistants comparable to
  Gemini Live~\cite{gemini_live_web_2026} and GPT-Live~\cite{openai_gpt_live_2026}, and was selected because several BLV participants already used it
  and it was freely available during recruitment.
  
% ── Tasks and Scenarios ─────────────────────────────────────────────────
\subsubsection{Tasks and Scenarios}
Informed by our formative study's three dominant challenge domains, we selected two representative tasks:
$\bullet$~\textbf{T1: Navigation (6 conditions).}
Participants used each system to navigate to a predefined destination.
Paths varied across two orthogonal dimensions: (1) \textit{Environment
Type}---\textit{Indoor}, \textit{Outdoor}, and \textit{Hybrid}; and
(2) \textit{Scene Complexity}---\textit{Low Clutter} (straightforward
route, minimal pedestrian traffic) or \textit{High Clutter} (convoluted
route, dense visual background, active pedestrian traffic).

$\bullet$~\textbf{T2: Object Retrieval (3 conditions).}
Participants located, identified, and grasped a specific target item among
distractors. Target Height was manipulated at three levels: \textit{High}
(above eye level), \textit{Medium} (at eye level), and \textit{Low}
(below eye level).

% ── Evaluation Metrics ──────────────────────────────────────────────────
\subsubsection{Evaluation Metrics}
Using a three-category battery:

$\bullet$~\textbf{Perceived Cognitive Load.} NASA Task Load Index (NASA-TLX)~\cite{hart2006nasa} assessed six dimensions (\textit{Mental Demand}, \textit{Physical Demand}, \textit{Temporal Demand}, \textit{Performance}, \textit{Effort}, \textit{Frustration}) on a 0--100 scale (5-point increments). We used the Raw TLX variant (unweighted) for comparable sensitivity and reduced burden~\cite{hart2006nasa}.

$\bullet$~\textbf{User Experience.} The System Usability Scale (SUS)~\cite{lewis2018system, brooke1996sus} comprises 10 items on a 5-point Likert scale (1=Strongly Disagree, 5=Strongly Agree); Brooke's procedure~\cite{brooke1996sus} converted responses to a 0--100 score. Items were verbally administered for accessibility.

$\bullet$~\textbf{Trust.} Single-item 7-point Likert measure (1=Not at all, 7=Completely): ``To what degree did you trust the assistant's information and suggestions?'' This isolated metric tests whether \textit{Calibrated Honesty} fosters higher trust than overconfident output.
% ── Procedure ───────────────────────────────────────────────────────────
\subsubsection{Procedure}
We conducted a within-subjects study where each participant evaluated all three systems. Each session lasted roughly 120\,minutes. To mitigate order effects, we incorporated two levels of counterbalancing: system presentation order followed a $3 \times 3$ Latin Square design~\cite{DBLP:conf/chi/ChenJMC025}, and within each system's trial block, the 9 tasks (6~navigation, 3~object retrieval) were randomised per participant. The procedure began with an onboarding phase for introductions, informed consent, and hands-on training for all systems. For each system, participants completed all 9 tasks, then immediately filled out the NASA-TLX, SUS, and trust measures for that system. For accessibility, a researcher administered these questionnaires verbally and recorded responses. After completing all three systems, the session concluded with a semi-structured interview to gather qualitative feedback comparing their overall experiences.

\subsection{Technical Evaluation}
\label{sec:tech_eval}
To provide complementary objective evidence, we conducted a technical evaluation on the full corpus of 243 user study trials, assessing task performance, output quality, and system latency.

% ── Task Performance ─────────────────────────────────────────────────────
\subsubsection{Task Performance}
We evaluated system efficiency using three objective metrics, each automatically logged per trial:

$\bullet$~\textbf{Task Completion Time.} Seconds from task initiation to experimenter-confirmed completion. Trials over 240\,s (navigation) or 180\,s (retrieval) were capped and marked as failures.

$\bullet$~\textbf{Conversational Turns.} Total number of user-system conversational turns recorded during the execution of each trial.

$\bullet$~\textbf{Task Success Rate.} Percentage of trials where the participant reached the target within the time limit, verified by an experimenter.

% ── Output Quality ───────────────────────────────────────────────────────
\subsubsection{Output Quality}
Two trained annotators independently coded all conversational transcripts (inter-rater $\kappa = 0.82$):

$\bullet$~\textbf{Navigation Accuracy.} Proportion of directional instructions aligned with experimenter-annotated ground-truth optimal paths.

$\bullet$~\textbf{Object Identification Accuracy.} Proportion of responses correctly identifying target, verified against experimenter annotations.

$\bullet$~\textbf{Sighted-Default Bias Rate.} Proportion of responses containing
  system-initiated visual-centric language that was not actionable for the user's task, such as ``look at,'' ``you can see,'' or color-only
  references. We treat this as the most direct and objectively measurable
  transcript-level signal of sighted-default bias, while recognizing that it
  captures only one dimension of the construct. The broader construct of
  sighted-default bias, including latency and output length, is characterized
  collectively by our metric suite.
  
% ── System Latency ───────────────────────────────────────────────────────
\subsubsection{System Latency}
End-to-end response latency was measured from user utterance offset to the onset of the system's audio response, logged using a system timer. We report  $p_{50}$, $p_{95}$, and $p_{99}$ percentiles across trials to characterize typical and tail-case performance, following standard practice for latency-sensitive systems.

\subsection{Data Analysis}
\label{sec:data_analysis}

We analyzed all data with a significance level of $\alpha = .05$. Our quantitative analysis followed a two-step process: we first ran an omnibus test for overall differences, and if the result was significant, we proceeded to pairwise comparisons. Pairwise $p$-values were adjusted using the Holm--Bonferroni procedure to control the family-wise error rate within each set of comparisons. For continuous metrics (Task Completion Time, Conversational Turns, accuracy rates), the omnibus test was a repeated-measures ANOVA followed by pairwise $t$-tests. For the binary Task Success Rate, we used Cochran's Q test followed by McNemar tests. All ordinal questionnaire data (NASA-TLX, SUS, Trust) were analyzed using Friedman tests followed by Wilcoxon signed-rank tests. Interview transcripts underwent reflexive thematic analysis~\cite{braun2006using}: two researchers independently coded transcripts, then iteratively converged on a shared codebook.
\section{Results}
\label{sec:results}

Our evaluation provides converging evidence that VIA-Agent's co-optimised architecture delivers significant improvements across objective performance, subjective experience, and component-level validation. We report results by evaluation method: technical evaluation (\S\ref{sec:tech_results}), user evaluation (\S\ref{sec:user_results}), and ablation study (\S\ref{sec:ablation_results}).

\subsection{Technical Evaluation Results}
\label{sec:tech_results}
\begin{table*}[t!]
  \centering
  \caption{Task completion times and statistical comparisons. We conducted a repeated-measures ANOVA (RM-ANOVA) for the omnibus test, reporting $F$, $p$, and partial eta-squared ($\eta_p^2$). For significant results, we performed post-hoc pairwise t-tests with Holm-Bonferroni correction, reporting adjusted $p$-values and Hedges' $g$ for effect size. Values of 240.00 indicate task timeout.}
  \label{tab:completion_time}
  \resizebox{\textwidth}{!}{%
\begin{tabular}{@{}ll ccc rrr *{3}{rr}@{}}
    \toprule
    % --- 第一行表头：逻辑分组 ---
    \multirow{2}{*}{\textbf{Task}} & \multirow{2}{*}{\textbf{Condition}} & \multicolumn{3}{c}{\textbf{Completion Time (s) $\downarrow$ (M $\pm$ SD)}} & \multicolumn{3}{c}{\textbf{Omnibus Test (RM-ANOVA)}} & \multicolumn{2}{c}{\textbf{VIA vs. Doubao}} & \multicolumn{2}{c}{\textbf{VIA vs. BeMyAI}} & \multicolumn{2}{c}{\textbf{Doubao vs. BeMyAI}} \\
    \cmidrule(lr){3-5} \cmidrule(lr){6-8} \cmidrule(lr){9-10} \cmidrule(lr){11-12} \cmidrule(lr){13-14}
    
    % --- 第二行表头：具体指标 ---
    & & {\textbf{VIA-Agent}} & {\textbf{Doubao}} & {\textbf{BeMyAI}} & {\textbf{$F$-stat}} & {\textbf{$p$-val}} & {\textbf{$\eta_p^2$}} & {\textbf{$p_{adj}$}} & {\textbf{$g$}} & {\textbf{$p_{adj}$}} & {\textbf{$g$}} & {\textbf{$p_{adj}$}} & {\textbf{$g$}} \\
    \midrule
    
    % --- Navigation 数据部分 ---
    \multirow{6}{*}{\textbf{Navigation}}
    & Indoor, Low Clutter    & $\textbf{66.00} \pm 27.16$ & $94.22 \pm 56.27$ & $240.00 \pm 0.00$ & 92.6194 & 0.0000 & 0.8340 & \textbf{0.0390} & 0.6083 & 0.0000 & 8.6282 & 0.0001 & 3.4891 \\
    & Indoor, High Clutter   & $\textbf{96.11} \pm 33.98$ & $159.89 \pm 70.10$ & $240.00 \pm 0.00$ & 24.9784 & 0.0000 & 0.6584 & \textbf{0.0288} & 1.1027 & 0.0000 & 5.7034 & 0.0179 &  1.5392 \\
    & Outdoor, Low Clutter   & $\textbf{56.33} \pm 13.83$ & $67.44 \pm 17.52$ & $240.00 \pm 0.00$ & 1080.2658 & 0.0000 &  0.9795 & \textbf{0.0004} &  0.6704 & 0.0000 & 17.8877 & 0.0000 & 13.2637 \\
    & Outdoor, High Clutter  & $\textbf{131.78} \pm 52.53$ & $163.56 \pm 66.41$ & $240.00 \pm 0.00$ & 19.3098 & 0.0001 & 0.4927 & \textbf{0.0384} & 0.5055 & 0.0008 & 2.7749 & 0.0173 & 1.5503 \\
    & Hybrid, Low Clutter    & $\textbf{82.11} \pm 20.33$ & $103.22 \pm 24.85$ & $240.00 \pm 0.00$ & 331.8882 & 0.0000 & 0.9413 & \textbf{0.0011} & 0.8856 & 0.0000 & 10.4595 & 0.0000 & 7.4138 \\
    & Hybrid, High Clutter   & $\textbf{191.11} \pm 46.30$ & $216.22 \pm 44.44$ & $240.00 \pm 0.00$ & 7.1382 & 0.0061 & 0.2461 & \textbf{0.0167} & 0.5270 & 0.0265 & 1.4220 & 0.1472 & 0.7206 \\
    
    \addlinespace 
    \midrule

    % --- Object Retrieval 数据部分 ---
    \multirow{3}{*}{\shortstack{\textbf{Object} \\ \textbf{Retrieval}}}
    & High Level             & $\textbf{68.44} \pm 22.40$ & $93.89 \pm 40.12$ & $162.89 \pm 23.54$ & 48.2826 & 0.0000 & 0.6684 & \textbf{0.0280} & 0.7459 & 0.0000 & 3.9147 & 0.0008 & 1.9980 \\
    & Medium Level           & $\textbf{36.00} \pm 9.07$ & $46.78 \pm 9.60$ & $130.56 \pm 42.53$ & 48.1961 & 0.0000 & 0.7525 & \textbf{0.0000} & 1.0991 & 0.0001 & 2.9286 & 0.0002 & 2.5880 \\
    & Low Level              & $\textbf{90.78} \pm 59.66$ & $105.22 \pm 56.15$ & $167.78 \pm 27.32$ & 17.7253 & 0.0001 & 0.3357 & \textbf{0.0081} & 0.2375 & 0.0068 & 1.5806 & 0.0081 & 1.3494 \\
    
    \bottomrule
  \end{tabular}%
  }
\end{table*}
\begin{table*}[t!]
  \centering
  \caption{Conversational turns and statistical comparisons. For each condition, we conducted an omnibus repeated-measures ANOVA (RM-ANOVA), reporting $F$, $p$, and partial eta-squared ($\eta_p^2$). For significant results, we performed Holm-Bonferroni corrected pairwise t-tests, reporting adjusted $p$-values and Hedges' $g$ for effect size. Non-significant comparisons are marked `/'.}
  \label{tab:conversational_turns}
  \resizebox{\textwidth}{!}{%
\begin{tabular}{@{}ll ccc rrr *{3}{rr}@{}}
    \toprule
    % --- 第一行表头 ---
    \multirow{2}{*}{\textbf{Task}} & \multirow{2}{*}{\textbf{Condition}} & \multicolumn{3}{c}{\textbf{Conversational Turns $\downarrow$ (M $\pm$ SD)}} & \multicolumn{3}{c}{\textbf{Omnibus Test (RM-ANOVA)}} & \multicolumn{2}{c}{\textbf{VIA vs. Doubao}} & \multicolumn{2}{c}{\textbf{VIA vs. BeMyAI}} & \multicolumn{2}{c}{\textbf{Doubao vs. BeMyAI}} \\
    \cmidrule(lr){3-5} \cmidrule(lr){6-8} \cmidrule(lr){9-10} \cmidrule(lr){11-12} \cmidrule(lr){13-14}
    
    % --- 第二行表头 ---
    & & {\textbf{VIA-Agent}} & {\textbf{Doubao}} & {\textbf{BeMyAI}} & {\textbf{$F$-stat}} & {\textbf{$p$-val}} & {\textbf{$\eta_p^2$}} & {\textbf{$p_{adj}$}} & {\textbf{$g$}} & {\textbf{$p_{adj}$}} & {\textbf{$g$}} & {\textbf{$p_{adj}$}} & {\textbf{$g$}} \\
    \midrule
    
    % --- Navigation 数据部分 ---
    \multirow{6}{*}{\textbf{Navigation}}
    & Indoor, Low Clutter    & $\textbf{3.78} \pm 1.56$ & $4.89 \pm 2.52$ & $4.11 \pm 0.93$ & 1.7035 & 0.2134 & 0.0703 & {/} & {/} & {/} & {/} & {/} & {/} \\
    & Indoor, High Clutter   & $\textbf{4.89} \pm 2.09$ & $8.22 \pm 2.95$ & $5.11 \pm 1.54$ & 9.0147 & 0.0024 & 0.3364 & \textbf{0.0312} & 1.2425 & 0.7287 & 0.1154 & 0.0312 & 1.2602 \\
    & Outdoor, Low Clutter   & $\textbf{3.67} \pm 1.12$ & $4.89 \pm 1.05$ & $5.11 \pm 1.76$ & 8.9091 & 0.0025 & 0.1992 & \textbf{0.0069} & 1.0713 & 0.0160 & 0.9316 & 0.5943 & 0.1457 \\
    & Outdoor, High Clutter  & $5.67 \pm 2.06$ & $7.56 \pm 2.83$ & $\textbf{4.44} \pm 0.88$ & 10.4737 & 0.0012 &  0.2975 & \textbf{0.0096} & 0.7261 & 0.0836 & 0.7342 & 0.0176 & 1.4121 \\
    & Hybrid, Low Clutter    & $\textbf{4.11} \pm 1.05$ & $5.89 \pm 1.27$ & $\textbf{4.11} \pm 0.78$ & 23.8140 & 0.0000 & 0.4156 & \textbf{0.0000} & 1.4512 & 1.0000 & 0.0000 & 0.0024 & 1.6062 \\
    & Hybrid, High Clutter   & $7.22 \pm 1.64$ & $8.78 \pm 1.79$ & $\textbf{4.33} \pm 0.87$ & 27.0164 & 0.0000 & 0.6329 & \textbf{0.0015} & 0.8634 & 0.0032 & 2.0965 & 0.0010 & 3.0140 \\
    
    \addlinespace 
    \midrule

    % --- Object Retrieval 数据部分---
    \multirow{3}{*}{\shortstack{\textbf{Object} \\ \textbf{Retrieval}}}
    & High Level             & $\textbf{3.22} \pm 0.97$ & $4.78 \pm 1.56$ & $3.56 \pm 0.53$ & 8.0994 & 0.0037 & 0.2916 & \textbf{0.0052} & 1.1381 & 0.3972 & 0.4061 & 0.0768 & 0.9977 \\
    & Medium Level           & $\textbf{2.22} \pm 0.67$ & $3.11 \pm 0.60$ & $3.33 \pm 0.87$ & 7.0000 & 0.0065 & 0.3333 & \textbf{0.0065} & 1.3339 & 0.0425 & 1.3693 & 0.5121 & 0.2839 \\
    & Low Level              & $3.78 \pm 1.86$ & $4.89 \pm 1.69$ & $\textbf{3.56} \pm 0.73$ & 3.6471 & 0.0495 & 0.1439 & \textbf{0.0082} & 0.5960 & 0.7458 & 0.1502 & 0.0995 & 0.9755 \\
    
    \bottomrule
  \end{tabular}%
  }
\end{table*}
\begin{table*}[t!]
  \centering
  \caption{Success rates and statistical comparisons. We conducted a Cochran's Q test for each condition, reporting $\chi^2$, $p$, and Cohen's $w$ for the omnibus effect size. For significant omnibus results, post-hoc pairwise McNemar's tests with Holm-Bonferroni correction were performed, reporting adjusted $p$ values and Cohen's $g$ for effect size.}
  \label{tab:success_rates}
  \resizebox{\textwidth}{!}{%
\begin{tabular}{@{}ll rrr rrr *{3}{rr}@{}}
    \toprule
    \multirow{2}{*}{\textbf{Task}} & \multirow{2}{*}{\textbf{Condition}} & \multicolumn{3}{c}{\textbf{Success Rate (\%) $\uparrow$}} & \multicolumn{3}{c}{\textbf{Omnibus Test (Cochran's Q)}} & \multicolumn{2}{c}{\textbf{VIA vs. Doubao}} & \multicolumn{2}{c}{\textbf{VIA vs. BeMyAI}} & \multicolumn{2}{c}{\textbf{Doubao vs. BeMyAI}} \\
    \cmidrule(lr){3-5} \cmidrule(lr){6-8} \cmidrule(lr){9-10} \cmidrule(lr){11-12} \cmidrule(lr){13-14}
    
    & & {\textbf{VIA-Agent}} & {\textbf{Doubao}} & {\textbf{BeMyAI}} & {\textbf{$\chi^2$}} & {\textbf{$p$-val}} & {\textbf{$w$}} & {\textbf{$p_{adj}$}} & {\textbf{$g$}} & {\textbf{$p_{adj}$}} & {\textbf{$g$}} & {\textbf{$p_{adj}$}} & {\textbf{$g$}} \\
    \midrule
    
    % --- Navigation 数据部分 (g 值建议用脚本计算) ---
    \multirow{6}{*}{\textbf{Navigation}}
    & Indoor, Low Clutter    & \textbf{100.0} & 88.9 & 0.0 & 16.22 & 0.0003 & 0.7751 & 1.0000 & 0.5000 & 0.0117 & 0.5000 & 0.0156 & 0.5000 \\
    & Indoor, High Clutter   & \textbf{88.9}  & 66.7 & 0.0 & 11.56 & 0.0031 & 0.6542 & 0.6250 & 0.2500 & 0.0234 & 0.5000 & 0.0625 & 0.5000 \\
    & Outdoor, Low Clutter   & \textbf{100.0} & \textbf{100.0} & 0.0 & 18.00 & 0.0001 & 0.8165 & 1.0000 & 0.0000 & 0.0117 & 0.5000 & 0.0117 & 0.5000 \\
    & Outdoor, High Clutter  & \textbf{88.9}  & 66.7 & 0.0 & 13.00 & 0.0015 & 0.6939 & 0.5000 & 0.5000 & 0.0234 & 0.5000 & 0.0625 & 0.5000 \\
    & Hybrid, Low Clutter    & \textbf{88.9}  & \textbf{88.9}  & 0.0 & 14.22 & 0.0008 & 0.7258 & 1.0000 & 0.0000 & 0.0234 & 0.5000 & 0.0234 & 0.5000 \\
    & Hybrid, High Clutter   & \textbf{55.6}  & 44.4 & 0.0 & 8.40  & 0.0150 & 0.5578 & 1.0000 &  0.5000 &  0.1875 & 0.5000 & 0.2500 & 0.5000 \\
    
    \addlinespace 
    \midrule

    % --- Object Retrieval 数据部分 ---
    \multirow{3}{*}{\shortstack{\textbf{Object} \\ \textbf{Retrieval}}}
    & High Level             & \textbf{100.0} & 88.9 & 44.4 & 8.40 & 0.0150 & 0.5578 & 1.0000 & 0.5000 & 0.1875 & 0.5000 & 0.2500 & 0.5000 \\
    & Medium Level           & \textbf{100.0} & \textbf{100.0} & 66.7 & 6.00 & 0.0498 & 0.4714 & 1.0000 & 0.0000 & 0.7500 & 0.5000 & 0.7500 & 0.5000 \\
    & Low Level              & \textbf{77.8}  & \textbf{77.8}  & 33.3 & 8.00 & 0.0183 & 0.5443 & 1.0000 & 0.0000 & 0.3750 & 0.5000 & 0.3750 & 0.5000 \\
    
    \bottomrule
  \end{tabular}%
  }
\end{table*}
% --- 5.1 Task Efficiency ------------------------------------------------------
\subsubsection{Task Performance} Evaluation yielded the following results:

$~\bullet$~\textbf{Task Completion Time.} VIA-Agent was significantly faster across all conditions (Table~\ref{tab:completion_time}). 
In navigation, VIA-Agent ($M=105.0$\,s) was 21.7\% faster than Doubao ($M=134.1$\,s) and substantially faster than BeMyAI, which timed out ($>240$\,s) in nearly all trials. 
In object retrieval, VIA-Agent ($M=65.1$\,s) was 20.6\% faster than Doubao ($M=82.0$\,s) and 57.6\% faster than BeMyAI ($M=153.7$\,s). 
Repeated-measures ANOVAs confirmed significant system effects ($p < .05, \eta_p^2 = .25$--$.98$); \textbf{post-hoc pairwise comparisons further confirmed that VIA-Agent significantly outperformed Doubao in all tested conditions (all $p_{adj} < .05$).}

$~\bullet$~\textbf{Conversational Turns.} VIA-Agent significantly reduced interaction overhead, requiring fewer turns than Doubao in eight of nine conditions (Table~\ref{tab:conversational_turns}). 
On average, VIA-Agent required 4.3 turns per trial compared to 5.9 for Doubao; \textbf{post-hoc tests confirmed VIA-Agent significantly outperformed Doubao in all conditions where a significant effect was found (all $p_{adj} < .05$).} 
This efficiency is attributed to VIA-Agent's goal-persistent architecture, which eliminates repetitive context clarifications. 
While BeMyAI occasionally showed fewer turns, this reflects significant processing latency per turn rather than true interaction efficiency.

$~\bullet$~\textbf{Task Success Rate.} VIA-Agent and Doubao both demonstrated high feasibility for assisting visually impaired users, significantly outperforming BeMyAI across most conditions (Table~\ref{tab:success_rates}). While VIA-Agent’s overall success rate ($88.9\%$) equaled or exceeded Doubao’s ($80.3\%$) in all nine conditions, this numerical advantage did not reach statistical significance due to ceiling effects. Crucially, however, \textbf{VIA-Agent achieved this high reliability with significantly greater efficiency}, requiring substantially fewer conversational turns and less completion time than Doubao, thereby proving its superior practical utility in real-time guidance.

% ── Output Quality───────────────────────────────────────────────────
\subsubsection{Output Quality}
\begin{table*}[t!]
  \centering
  \caption{Output quality metrics and statistical comparisons. We conducted repeated-measures ANOVA (RM-ANOVA) for omnibus tests, reporting $F$, $p$, and partial eta-squared ($\eta_p^2$). For significant results, we performed post-hoc pairwise t-tests with Holm-Bonferroni correction, reporting adjusted $p$-values and Hedges' $g$ for effect size.}
  \label{tab:output_quality}
  \resizebox{\textwidth}{!}{%
\begin{tabular}{@{}l ccc rrr *{3}{rr}@{}}
    \toprule
    \multirow{2}{*}{\textbf{Metric}} & \multicolumn{3}{c}{\textbf{Score (\%) (M $\pm$ SD)}} & \multicolumn{3}{c}{\textbf{Omnibus (RM-ANOVA)}} & \multicolumn{2}{c}{\textbf{VIA vs. Doubao}} & \multicolumn{2}{c}{\textbf{VIA vs. BeMyAI}} & \multicolumn{2}{c}{\textbf{Doubao vs. BeMyAI}} \\
    \cmidrule(lr){2-4} \cmidrule(lr){5-7} \cmidrule(lr){8-9} \cmidrule(lr){10-11} \cmidrule(lr){12-13}
    & {\textbf{VIA-Agent}} & {\textbf{Doubao}} & {\textbf{BeMyAI}} & {\textbf{$F$}} & {\textbf{$p$}} & {\textbf{$\eta_p^2$}} & {\textbf{$p_{adj}$}} & {\textbf{$g$}} & {\textbf{$p_{adj}$}} & {\textbf{$g$}} & {\textbf{$p_{adj}$}} & {\textbf{$g$}} \\
    \midrule
    Navigation Accuracy (\%) $\uparrow$ & $\textbf{80.1} \pm 2.3$ & $64.5 \pm 5.0$ & $28.6 \pm 3.5$ & 402.37 & 0.0000 & 0.9732 & \textbf{0.0001} & 3.8065 & 0.0000 & 16.3674 & 0.0000 & 7.8552 \\
    Object Identification Accuracy (\%)  $\uparrow$ & $\textbf{94.2} \pm 0.8$ & $88.7 \pm 2.4$ & $30.8 \pm 5.9$ & 724.95 & 0.0000 & 0.9852 & \textbf{0.0003} & 2.9157 & 0.0000 & 14.2372 & 0.0000 & 12.1685 \\
    Sighted-Default Bias Rate (\%) $\downarrow$ & $\textbf{31.3} \pm 2.8$ & $89.8 \pm 0.8$ & $81.8 \pm 4.9$ & 1027.48 & 0.0000 & 0.9860 & \textbf{0.0000} & 26.7978 & 0.0000 & 12.1109 & 0.0016 & 2.1748 \\ 
    \bottomrule
  \end{tabular}%
  }
\end{table*}
Table~\ref{tab:output_quality} summarizes the results:

$~\bullet$~\textbf{Navigation Accuracy.} VIA-Agent achieved significantly higher navigation accuracy ($M = 80.1\%$) than both Doubao ($M = 64.5\%$) and BeMyAI ($M = 28.6\%$). Doubao also outperformed BeMyAI. The accuracy advantage reflects VIA-Agent's goal-persistent reasoning. Notably, Doubao frequently hallucinated by associating environmental cues with online resources rather than grounding responses in the physical destination, causing misdirected instructions.

$~\bullet$~\textbf{Object Identification Accuracy.} Similarly, VIA-Agent achieved the highest accuracy ($M = 94.2\%$), significantly outperforming Doubao ($M = 88.7\%$) and BeMyAI ($M = 30.8\%$). Doubao significantly outperformed BeMyAI. A similar failure pattern was observed with Doubao, which occasionally referenced online resources or general knowledge rather than identifying the physical target object, reducing its grounded identification reliability.

$~\bullet$~\textbf{Sighted-Default Bias Rate.}
VIA-Agent exhibited significantly lower bias rates ($M = 31.3\%$) than both Doubao ($M = 89.8\%$) and BeMyAI ($M = 81.8\%$). Notably, both baselines produced vision-centric language in over 80\% of responses, confirming pervasive sighted-default assumptions in current VLMs---even BeMyAI, explicitly designed for BLV users, exhibited substantial bias. VIA-Agent's substantially lower rate validates the Calibrated Honesty principle's effectiveness in reducing inaccessible visual-centric language and delivering more actionable, grounded information for BLV users.

% ── System Latency ───────────────────────────────────────────────────
\subsubsection{System Latency}
VIA-Agent ($p_{50}$=2.8s, $p_{95}$=7.9s, $p_{99}$=14.2s) achieved comparable per-response latency to Doubao ($p_{50}$=2.3s, $p_{95}$=7.6s, $p_{99}$=11.7s), with modest overhead from the bias-filtering pipeline. Both systems exhibited a significant speed lead over BeMyAI ($p_{50}$=7.4s, $p_{95}$=13.9s, $p_{99}$=20.5s), fulfilling the Continuity principle.
This latency profile should be interpreted by distinguishing network latency from end-to-end response latency. The communication between the VLM and the user is real-time: our RTC layer (based on veRTC~\cite{volcengine_vertc_web_2026}) has a network latency of 100--200~ms. The $p_{95}$=7.9s / $p_{99}$=14.2s tail originates from VLM inference latency, an intrinsic cost of thinking-enabled models. Accordingly, our use of ``real-time'' refers to continuous streaming and task-level interaction efficiency rather than sub-second obstacle avoidance.
Despite slightly higher per-response latency, VIA-Agent achieved faster task completion (Table~\ref{tab:completion_time}) and significantly fewer conversational turns (Table~\ref{tab:conversational_turns}), validating the Conciseness principle.

\subsection{User Evaluation Results}
\label{sec:user_results}
  \begin{figure*}[t!]                                                   
    \centering            
    \includegraphics[width=\textwidth]{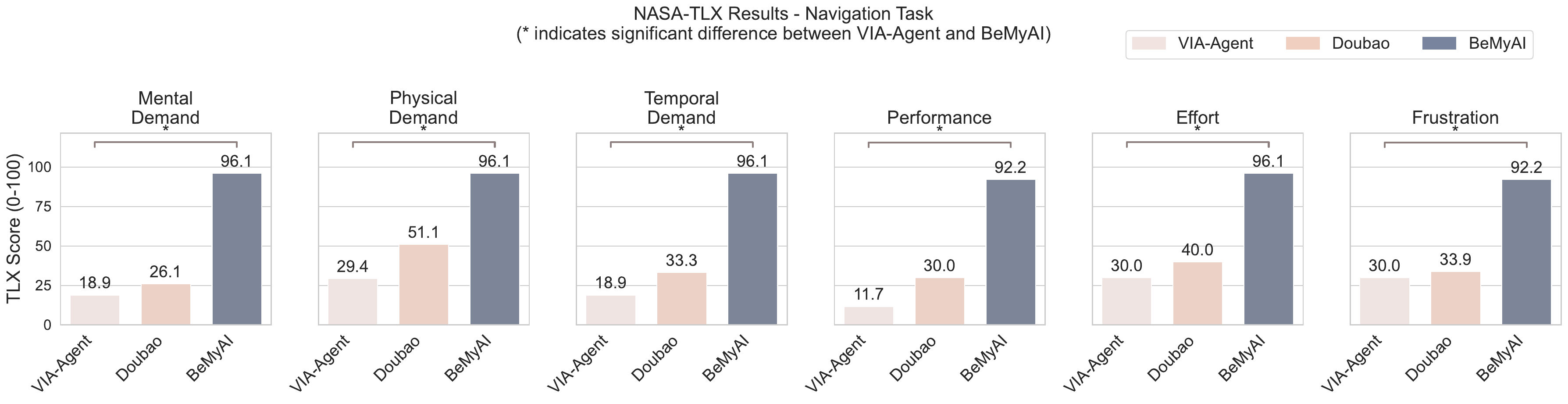}   
    \caption{Navigation NASA-TLX. Asterisks show significant VIA-Agent vs. BeMyAI differences (Holm-Bonferroni corrected).}
    \Description{Navigation NASA-TLX. Asterisks show significant VIA-Agent vs. BeMyAI differences (Holm-Bonferroni corrected).}  
    \label{fig:nav_tlx}                                                 
  \end{figure*}

  \begin{figure*}[t!]                                                  
    \centering
    \includegraphics[width=\textwidth]{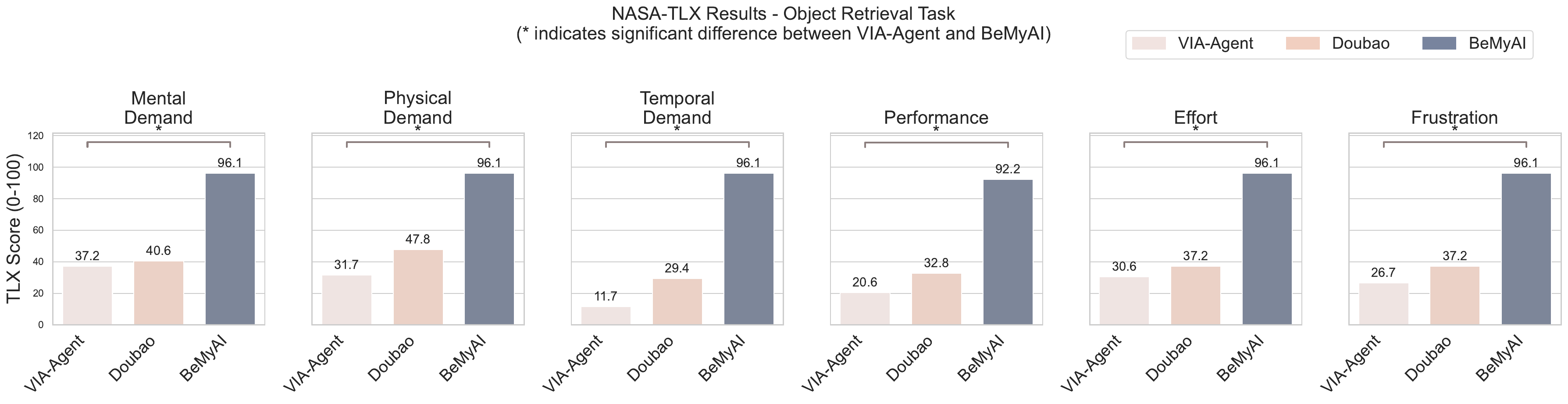}
    \caption{Object Retrieval NASA-TLX. Asterisks show significant VIA-Agent vs. BeMyAI differences (Holm-Bonferroni corrected).}
    \Description{Object Retrieval NASA-TLX. Asterisks show significant VIA-Agent vs. BeMyAI differences (Holm-Bonferroni corrected).}
    \label{fig:obj_tlx}
  \end{figure*}

% ── Perceived Cognitive Load ─────────────────────────────────────────
$~\bullet$~\textbf{Perceived Cognitive Load.}
VIA-Agent significantly outperformed BeMyAI across all six NASA-TLX dimensions (Figure~\ref{fig:nav_tlx}, ~\ref{fig:obj_tlx}), including lower effort and better performance. Qualitative data mirrored this, as users found BeMyAI’s manual image capture cumbersome and its verbose responses overwhelming. While VIA-Agent scored numerically better than the streaming baseline Doubao, differences were not statistically significant; users valued Doubao's fluid conversational style despite its lack of specialized features. Crucially, both streaming systems (VIA-Agent and Doubao) substantially outperformed the request-response model (BeMyAI), underscoring that seamless, real-time interaction is vital for reducing cognitive load in assistive vision tasks.

% ── User Experience ──────────────────────────────────────────────────
$~\bullet$~\textbf{User Experience.}
System Usability Scale (SUS) scores indicated significant differences in overall usability among the three systems. VIA-Agent achieved the highest mean score ($M = 83.5$), outperforming both Doubao ($M = 78.4$) and BeMyAI ($M = 32.4, p < .05$). These findings align with the NASA-TLX results: the real-time, streaming architecture of VIA-Agent and Doubao effectively reduced the cognitive overhead associated with BeMyAI's static interface, resulting in higher user confidence and overall satisfaction.

% ── Trust ────────────────────────────────────────────────────────────
$~\bullet$~\textbf{Trust.}
Trust scores revealed significant differences across three systems. VIA-Agent ($M=6.22$) significantly outperformed both Doubao ($M=5.11; p=.016$) and BeMyAI ($M=3.33; p=.012$). These findings suggest that Calibrated Honesty improved perceived reliability, but trust alone does not establish calibration. Transcript-level analysis provides complementary evidence: in object identification, VIA-Agent reduced
  overconfident incorrect assertions from approximately 11\% for Doubao to
  approximately 3\%. 
  In addition, VIA-Agent's confident assertions were 96\% accurate, compared with 81\% for hedged assertions, indicating alignment between expressed confidence and correctness.
  Rather than presenting all generated claims as objective facts, the agent's ability to filter outputs based on internal confidence levels fosters user reliability and trust.

% ── Qualitative Findings ─────────────────────────────────────────────
$~\bullet$~\textbf{Qualitative Findings.}
Thematic analysis identified three key themes: (1) \textbf{Continuity as Threshold.} Participants unanimously preferred streaming systems over BeMyAI, citing reduced cognitive burden: \textit{``With VIA-Agent, I don't have to stop and think about when to take a photo. It just flows''} (P4). BeMyAI's manual capture workflow was described as \textit{``exhausting''} (P7) and \textit{``breaking my concentration''} (P2). (2) \textbf{Conciseness Reduces Overload.} Participants noted VIA-Agent's responses were \textit{``straight to the point''} (P6) compared to Doubao's \textit{``too much information at once''} (P3). One participant explained: \textit{``Doubao tells me everything it sees, but I just need to know: left or right?''} (P8). (3) \textbf{Honesty Builds Trust.} Participants appreciated VIA-Agent's uncertainty disclosure: \textit{``When it says 'I'm not sure,' I know to double-check. That's more helpful than pretending it knows''} (P5). This contrasts with Doubao's overconfident errors, which participants found \textit{``dangerous''} (P1) and \textit{``misleading''} (P9).

\subsection{Ablation Study}
\label{sec:ablation_results}
To isolate each design principle's contribution, we conducted an ablation study in the high-clutter indoor navigation scenario. Results confirm all three design principles are necessary (Table~\ref{tab:ablation}). The \textbf{Full VIA-Agent} significantly outperformed all variants across all metrics, achieving the shortest task time ($M = 96.11$\,s), most concise feedback ($M = 28.5$ words), and highest trust ($M = 6.22$). Specifically, our analysis reveals:
(1) \textbf{Continuity ($P_1$) as the Foundation:} Removing $P_1$ breaks real-time interaction and cannot be scored in Table~\ref{tab:ablation}: our $-P_1$ variant, the MCP embodiment (Sec.~\ref{sec: MCP_embodiment}), ran in request--response mode with $>$15\,s per-turn latency, making navigation tasks uncompletable. Its necessity is instead evidenced by this latency and by BeMyAI (also request--response) timing out in nearly all navigation trials (Tables~\ref{tab:completion_time}), establishing that a continuous feedback loop is a prerequisite for real-time assistance. (2) \textbf{Conciseness ($P_2$) for Efficiency:} Removing $P_2$ resulted in a 66.0\% increase in output verbosity ($M = 47.3$ vs. $28.5$ words) and a 37.8\% increase in task time ($M = 132.40$\,s). This demonstrates that concise instructions are key to reducing cognitive load during navigation. (3) \textbf{Calibrated Honesty ($P_3$) for Trust:} While having less impact on task time, removing $P_3$ significantly lowered user trust to $4.67$ ($SD = 0.87$). This validates $P_3$’s unique role in managing user expectations through uncertainty awareness. In summary, $P_1$, $P_2$, and $P_3$ independently and synergistically address distinct challenges in AI assistance for BLV users.
%--------------------------------------------------------------------
\begin{table}[t]\centering
  \small
  \caption{Ablation results (high-clutter indoor navigation). Each variant removes exactly one component, holding all else fixed. Time in seconds, output length in words, trust on a 1--7 scale. Values are mean\,$\pm$\,SD. *$p < 0.05$ vs.\ Full VIA-Agent.}
  \label{tab:ablation}
  \setlength{\tabcolsep}{3pt}
  \renewcommand{\arraystretch}{1.2}
  \begin{tabular}{@{}p{64pt} r@{\,$\pm$\,}l r@{\,$\pm$\,}l r@{\,$\pm$\,}l@{}}
  \toprule
  \textbf{Condition} & \multicolumn{2}{c}{\textbf{Time (s)}} & \multicolumn{2}{c}{\textbf{Words}} & \multicolumn{2}{c}{\textbf{Trust}} \\
  \midrule
  Full VIA-Agent & 96.11 & 33.98 & 28.5 & 5.2 & 6.22 & 0.67 \\
  \midrule
  $-P_2$ (w/o Conciseness) & 132.40 & 45.17\rlap{*} & 47.3 & 6.2\rlap{*} & 5.89 & 0.78 \\
  $-P_3$ (w/o Calibrated Honesty) & 127.19 & 41.82\rlap{*} & 34.6 & 5.7 & 4.67 & 0.87\rlap{*} \\
  \bottomrule
  \end{tabular}
\end{table}

\section{Discussion}
\label{sec:discussion}
Our evaluation demonstrates that VIA-Agent’s design principles effectively mitigate the \textit{sighted-default bias} inherent in general-purpose VLMs. The 21.4\% reduction in task time and 27\% fewer conversational turns compared to Doubao signify more than mere efficiency gains; they represent a fundamental architectural shift from \textbf{visual-centric narration} to \textbf{auditory-first, goal-directed guidance}. While baseline systems like Doubao provide high-fidelity descriptions, they implicitly assume a user capable of parallel visual processing—filtering through verbose spatial data to extract navigation cues. In contrast, VIA-Agent’s performance validates that for BLV users, information overload is often a byproduct of \textit{inappropriate delivery} rather than volume. By prioritizing serialized auditory perception, VIA-Agent transforms the user from a passive listener filtering environmental noise into an active decision-maker responding to streamlined signals.

The trust results suggest that Calibrated Honesty is not a compromise but a necessity. Participants rated VIA-Agent significantly more trustworthy than Doubao precisely because its multi-level confidence filtering avoids the presentation of overconfident, unverifiable claims~\cite{chang2025probing-gaps-chatgpt-live}. For BLV users who cannot visually verify outputs, epistemic honesty is a fundamental safety requirement.

\textbf{Limitations.} Our modest sample size ($N=9$) is consistent with comparable BLV evaluations~\cite{killough2025vrsight,huh2025vid2coach}. VIA-Agent's smartphone implementation may limit hands-free usability, a necessary trade-off given current hardware constraints for low-latency RTC streaming.

\textbf{Design Implications.} Our work yields three key implications for assistive AI: (1) perceptual modality must drive architectural decisions—low latency is insufficient without designing for serial auditory perception; (2) specialization outperforms generalization—purpose-built cognitive architectures are required to eliminate implicit sighted-default assumptions; (3) epistemic honesty is a safety requirement for users who cannot visually verify AI outputs.

\textbf{Future Work.} Key directions include extending the VIA-Agent framework to additional domains (social navigation, document analysis), developing wearable embodiments once hardware supports low-latency multimodal RTC, and conducting longitudinal studies to understand how trust evolves over extended real-world use.

\section{Conclusion}
\label{sec:conclusion}
In this work, we present VIA-Agent, a framework addressing the sighted-default bias by integrating Continuity, Conciseness, and Calibrated Honesty into real-time assistance. By co-optimizing a specialized cognitive core with a low-latency RTC architecture, VIA-Agent facilitates navigation and object retrieval in dynamic environments. Rigorous within-subjects evaluations validate its ability to significantly reduce cognitive load and enhance user trust.

\begin{acks}
This work was supported by the STIC Shenzhen Natural Science Foundation
(No.~JCYJ20250604191329039) and the Innovation and Technology Fund
(Project No.~PRP/047/22FX). We also thank the anonymous reviewers for their
constructive feedback.
\end{acks}
%% The next two lines define the bibliography style to be used, and
%% the bibliography file.
\bibliographystyle{ACM-Reference-Format}
\bibliography{sample-base}

\appendix
\section{VIA-Agent in Use}
\label{sec:appendix_in_use}
\begin{figure*}[t]
\centering
\includegraphics[width=\textwidth]{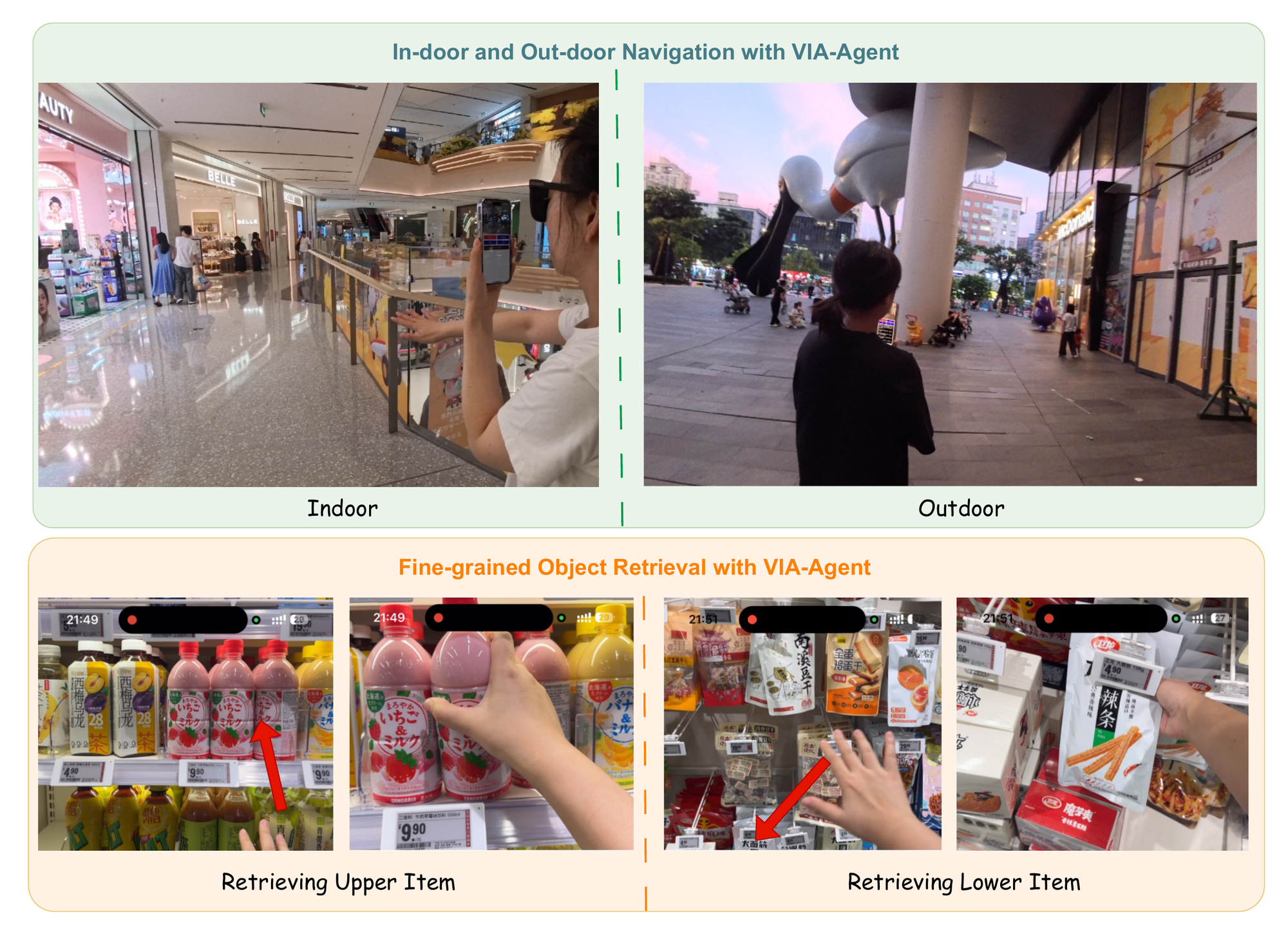}
\caption{Representative evaluation conditions. Top: indoor and outdoor
navigation scenarios. Bottom: object-retrieval tasks with targets placed at
different heights among visually similar distractors.}
\Description{Representative evaluation conditions. Top: indoor and outdoor
navigation scenarios. Bottom: object-retrieval tasks with targets placed at
different heights among visually similar distractors.}
\label{fig:evaluation_conditions}
\end{figure*}

Figure~\ref{fig:evaluation_conditions}
  shows the broader evaluation settings used in the study, including indoor and
  outdoor navigation scenes and object-retrieval tasks with targets placed at
  different heights among distractors.

\end{document}